\documentclass[journal]{IEEEtran}
\IEEEoverridecommandlockouts

\ifCLASSINFOpdf
   \usepackage{graphicx}
\else
   \usepackage[dvips]{graphicx}
\fi
\usepackage[cmex10]{amsmath}
\usepackage[utf8]{inputenc}
\usepackage{xcolor}
\usepackage{colortbl}
\usepackage{array}
\usepackage{eqparbox}
\usepackage{booktabs}

\usepackage{textcomp}
\usepackage{tikz}
\usetikzlibrary{positioning}
\usepackage{tikz-3dplot}
\usepackage{comment}
\usepackage{array}
\usepackage{eqparbox}
\usepackage{alphalph,etoolbox}
\patchcmd{\subequations}{\alph{equation}}{\alphalph{\value{equation}}}{}{}
\usepackage{multirow}
\usepackage{multicol}
\usepackage{booktabs}
\usepackage{graphics}
\usepackage{textcomp}
\usepackage{tikz}
\usetikzlibrary{shapes,arrows}
\usepackage{tikz-3dplot}
\usepackage{comment}
\usepackage{comment}
\usepackage{amsfonts}
\usepackage{amssymb}
\usepackage{amsthm}
\usepackage{cite}
\usepackage{algorithm}
\usepackage{algpseudocode}

\usepackage{pgfplots}
\usetikzlibrary{calc,shadings}

\newenvironment{customlegend}[1][]{%
    \begingroup
    \csname pgfplots@init@cleared@structures\endcsname
    \pgfplotsset{#1}%
}{%
    \csname pgfplots@createlegend\endcsname
    \endgroup
}%

\def\addlegendimage{\csname pgfplots@addlegendimage\endcsname}

\usepackage{nomencl}
\makenomenclature
\usepackage{etoolbox}
\usepackage[official]{eurosym}
\usepackage{booktabs}
\usepackage{multirow}
\usepackage{cancel}
\definecolor{Gray}{gray}{0.9}
\definecolor{LightCyan}{rgb}{0.88,1,1}
\usepackage{alphalph}

\AtBeginDocument{%
  \AtBeginEnvironment{subequations}{%
    \let\alph\alphalphval%
  }}
\usepackage{tikz}
\usepackage{pgfplots}
\pgfplotsset{compat=newest}
\usepgfplotslibrary{groupplots}
\usepgfplotslibrary{dateplot}
\usepackage{longtable}
\usepackage{empheq}
\usepackage{environ}
\makeatletter
\newsavebox{\measure@tikzpicture}
\NewEnviron{scaletikzpicturetowidth}[1]{%
	\def\tikz@width{#1}%
	\begin{lrbox}{\measure@tikzpicture}%
		\BODY
	\end{lrbox}%
	\pgfmathparse{#1/\wd\measure@tikzpicture}%
	\BODY
}
\newcommand{\rom}[1]{\lowercase\expandafter{\romannumeral #1\relax}}
\makeatother
\usepackage{svg}
\allowdisplaybreaks
\usepackage{longtable}
\usepackage{tabularx}
\newcolumntype{Y}{>{\centering\arraybackslash}X}
\usepackage{svg}
\usepackage{url}
\usepackage{hyperref}
\usepackage{tikz}
\usetikzlibrary{shapes,arrows}
\usepackage{tikz-3dplot}
\usepackage{pgfplots}
\usepackage[flushleft]{threeparttable}
\allowdisplaybreaks
\definecolor{color100}{rgb}{0.917647058823529,0.917647058823529,0.949019607843137}
\definecolor{color101}{rgb}{0.298039215686275,0.447058823529412,0.690196078431373}
\definecolor{color102}{rgb}{0.333333333333333,0.658823529411765,0.407843137254902}
\definecolor{color103}{rgb}{0.768627450980392,0.305882352941176,0.32156862745098}
\definecolor{color104}{rgb}{0.505882352941176,0.447058823529412,0.698039215686274}
\definecolor{color110}{rgb}{0.917647058823529,0.917647058823529,0.949019607843137}
\definecolor{color111}{rgb}{0.748039215686275,0.700980392156863,0.812745098039216}
\definecolor{color112}{rgb}{0.933823529411765,0.754411764705882,0.583823529411765}
\definecolor{color113}{rgb}{0.95,0.95,0.65}
\definecolor{color114}{rgb}{0.27843137254902,0.431372549019608,0.631372549019608}
\definecolor{color115}{rgb}{0.824509803921569,0.124509803921569,0.492156862745098}
\definecolor{color116}{rgb}{0.666666666666667,0.372549019607843,0.172549019607843}

\usepackage{scalerel}
\usetikzlibrary{svg.path}
\definecolor{orcidlogocol}{HTML}{A6CE39}
\tikzset{
orcidlogo/.pic={
\fill[orcidlogocol] svg{M256,128c0,70.7-57.3,128-128,128C57.3,256,0,198.7,0,128C0,57.3,57.3,0,128,0C198.7,0,256,57.3,256,128z};
\fill[white] svg{M86.3,186.2H70.9V79.1h15.4v48.4V186.2z}
svg{M108.9,79.1h41.6c39.6,0,57,28.3,57,53.6c0,27.5-21.5,53.6-56.8,53.6h-41.8V79.1z M124.3,172.4h24.5c34.9,0,42.9-26.5,42.9-39.7c0-21.5-13.7-39.7-43.7-39.7h-23.7V172.4z}
svg{M88.7,56.8c0,5.5-4.5,10.1-10.1,10.1c-5.6,0-10.1-4.6-10.1-10.1c0-5.6,4.5-10.1,10.1-10.1C84.2,46.7,88.7,51.3,88.7,56.8z};
}
}
\newcommand\orcidicon[1]{\href{https://orcid.org/#1}{\mbox{\scalerel*{
\begin{tikzpicture}[yscale=-1,transform shape]
\pic{orcidlogo};
\end{tikzpicture}
}{|}}}}

\usepackage{nomencl}
\makenomenclature
\renewcommand\nomgroup[1]{%
  \item[\bfseries
  \ifstrequal{#1}{I}{Indices}{%
      \ifstrequal{#1}{V}{Variables (lower-case)}{%
          \ifstrequal{#1}{C}{Constants}{%
              \ifstrequal{#1}{P}{Parameters (upper-case)}{%
                \ifstrequal{#1}{J}{Sets}{%
                \ifstrequal{#1}{O}{Operators and functions}{%
                }%
                }%
              }%
          }%
      }%
  }%
]}%
\newcommand{\p}{\textup{\texttt{+}}} 
\newcommand{\m}{\textup{\texttt{-}}} 
\newcommand{\lablstminput}{I}
\newcommand{\lablstmoutput}{O}
\newcommand{\lablstmskip}{S}

\newcommand{\labdcgansG}{G}
\newcommand{\labdcgansD}{D}

\newcommand{\setlstmT}{\mathbb{T}}
\newcommand{\indlstmt}{t}
\nomenclature[It]{$\indlstmt\in\setlstmT$}{Time index;}
\newcommand{\varlstmb}{b_{\indlstmt}}
\nomenclature[Vlstmb]{$\varlstmb$}{Bias in LSTM algorithm;}
\newcommand{\varlstmbi}{b^{\lablstminput}_{\indlstmt}}
\nomenclature[Vlstmbi]{$\varlstmbi$}{Input bias in LSTM algorithm;}
\newcommand{\varlstmbo}{b^{\lablstmoutput}_{\indlstmt}}
\nomenclature[Vlstmbo]{$\varlstmbo$}{Output bias in LSTM algorithm;}
\newcommand{\varlstmbs}{b^{\lablstmskip}_{\indlstmt}}
\nomenclature[Vlstmbs]{$\varlstmbs$}{Skip bias in LSTM algorithm;}
\newcommand{\varlstms}{s_{\indlstmt}}
\newcommand{\varlstmsminus}{s_{\indlstmt\m1}}
\nomenclature[Vlstms]{$\varlstms$}{State in LSTM algorithm;}
\newcommand{\varlstmx}{x_{\indlstmt}}
\nomenclature[Vlstmx]{$\varlstmx$}{Input in LSTM algorithm;}
\newcommand{\varlstmy}{y_{\indlstmt}}
\newcommand{\varlstmyminus}{y_{\indlstmt\m1}}
\newcommand{\varlstmyhat}{\hat{y}_{\indlstmt}}
\nomenclature[Vlstmy]{$\varlstmy$}{Output in LSTM algorithm;}
\newcommand{\vardcganslvec}{\boldsymbol{l}}
\nomenclature[Vdcgansl]{$\vardcganslvec$}{Latent space variable vector in DCGANs algorithm;}
\newcommand{\vardcgansxvec}{\boldsymbol{x}}
\nomenclature[Vdcgansx]{$\vardcgansxvec$}{Input variable vector in DCGANs algorithm;}
\newcommand{\vardcgansxhatvec}{\hat{\boldsymbol{x}}}
\nomenclature[Vdcgansxg]{$\vardcgansxhatvec$}{Generated input vector in DCGANs algorithm;}
\newcommand{\vardcgansyvec}{\boldsymbol{y}}
\nomenclature[Vdcgansy]{$\vardcgansyvec$}{Output variable vector in DCGANs algorithm;}
\newcommand{\parlstmw}{W}
\newcommand{\parlstmv}{V}
\nomenclature[Pw]{$\parlstmw$/$\parlstmv$}{Weights in LSTM algorithm;}
\newcommand{\parlstmwi}{W^{\lablstminput}}
\newcommand{\parlstmvi}{V^{\lablstminput}}
\nomenclature[Pwi]{$\parlstmwi$/$\parlstmvi$}{Weights for inputs in LSTM algorithm;}
\newcommand{\parlstmwo}{W^{\lablstmoutput}}
\newcommand{\parlstmvo}{V^{\lablstmoutput}}
\nomenclature[Pwo]{$\parlstmwo$/$\parlstmvo$}{Weights for outputs in LSTM algorithm;}
\newcommand{\parlstmws}{W^{\lablstmskip}}
\newcommand{\parlstmvs}{V^{\lablstmskip}}
\nomenclature[Pws]{$\parlstmws$/$\parlstmvs$}{Weights for skips in LSTM algorithm;}
\newcommand{\funlstmsigmoid}{\text{sigmoid}}
\newcommand{\funlstmtanh}{\text{tanh}}%
\newcommand{\funlstmprod}{\circ}
\nomenclature[Oh]{$\funlstmprod$}{Hadamard product of two matrices;}
\newcommand{\funlstmexpec}{\text{E}}
\newcommand{\fundcgansexpec}{\text{E}}
\nomenclature[Oe]{$\funlstmexpec(\cdot)$}{Expected value of ($\cdot$);}
\newcommand{\fundcgansnetG}{\text{N}^\labdcgansG}
\newcommand{\fundcgansnetD}{\text{N}^\labdcgansD}
\nomenclature[O]{$\fundcgansnetG$, $\fundcgansnetD$}{Network model for generator and discriminator in DCGANs algorithm;}

\begin{document}
\bstctlcite{IEEEexample:BSTcontrol}
\title{Econometric Modeling of Intraday Electricity Market Price with Inadequate Historical Data}
\author{
\IEEEauthorblockN{Saeed~Mohammadi\orcidicon{0000-0003-1823-9653} and Mohammad~Reza~Hesamzadeh\orcidicon{0000-0002-9998-9773},}
\IEEEauthorblockA{School of Electrical and Computer Engineering,\\
KTH Royal Institute of Technology,
Stockholm, Sweden, \{saeedmoh, mrhesamzadeh\}@kth.se
\vspace{-12mm}}
\thanks{\noindent This work was financially supported by the Swedish Energy Agency (Energimyndigheten) under Grant 3233. The required computation is performed by computing resources from the Swedish National Infrastructure for Computing (SNIC) at PDC center for high performance computing at KTH Royal Institute of Technology which was supported by the Swedish Research Council under Grant 2018-05973. (Corresponding author: Saeed Mohammadi.)}
}
\markboth{}{Saeed Mohammadi \MakeLowercase{\textit{et al.}}: Econometric Modeling of Intraday Electricity Market Price with Inadequate Historical Data}

\maketitle
\begin{abstract}
The intraday (ID) electricity market has received an increasing attention in the recent EU electricity-market discussions. This is partly because the uncertainty in the underlying power system is growing and the ID market provides an adjustment platform to deal with such uncertainties. Hence, market participants need a proper ID market price model to optimally adjust their positions by trading in the market. Inadequate historical data for ID market price makes it more challenging to model. This paper proposes long short-term memory, deep convolutional generative adversarial networks, and No-U-Turn sampler algorithms to model ID market prices. Our proposed econometric ID market price models are applied to the Nordic ID price data and their promising performance are illustrated.
\end{abstract}
\IEEEpeerreviewmaketitle
\vspace{-5mm}
\section{Introduction}\label{sec:introduction}
\IEEEPARstart{P}{urchasing} electricity for a specific bidding zone involves a considerable risk due to the price risk. One delicate arrangement is to go to a forward market, e.g. financial transmission right (FTR) or day-ahead (DA) market, and buy a forward contract \cite{Bjork2019arbitrage}. Then we would be able to guard ourselves against this risk with an option/binding contract for a specific delivery time. Although wholesale electricity markets operate as a multi-settlement system (i.e. DA and intraday (ID) electricity markets), the DA markets conventionally lead providing economic dispatch. To facilitate integration of renewable energy resources in these multi-settlement systems, ID markets play a significant rule in delivering electricity by offering the opportunity to trade power shortly before the start of delivery time. DA prices and renewable energy generation are not continually coherent due to lack of flexibility in DA markets. Therefore, liquid ID markets, energy storage systems, and demand flexibility are beneficial to avoid price peaks. This requires a reliable price modeling for these multi-settlement systems which is the main focus of this paper.
\vspace{-5mm}
\subsection{Motivation}\label{sec:motivation}
The ID market plays an increasing role in handling the power system uncertainties and accordingly facilitating integration of renewable energy resources. The ID market provides an adjustment platform to trade power shortly before start of the delivery time and hence liquid ID markets are beneficial to avoid real-time price spikes. Still, optimal trading in the ID market requires a proper ID market price modeling.

\begin{figure}[!h]
	\centering
\begin{tikzpicture}

\definecolor{color0}{rgb}{0.67843137254902,0.847058823529412,0.901960784313726}
\definecolor{color1}{rgb}{0.12156862745098,0.466666666666667,0.705882352941177}
\definecolor{color2}{rgb}{1,0.498039215686275,0.0549019607843137}
\definecolor{color3}{rgb}{0.172549019607843,0.627450980392157,0.172549019607843}
\definecolor{color4}{rgb}{0.83921568627451,0.152941176470588,0.156862745098039}

\begin{axis}[
legend cell align={left},
legend style={
  fill opacity=0.8,
  draw opacity=1,
  text opacity=1,
  at={(0.03,0.97)},
  anchor=north west,
  draw=white!80!black
},
tick align=outside,
tick pos=left,
title={\scriptsize DA/ID prices 2020-12-14-SE},
x grid style={white!69.0196078431373!black},
xlabel={\scriptsize Hour},
xmajorgrids,
xmin=-1.15, xmax=24.15,
xtick style={color=black},
xtick={1,3,5,7,9,11,13,15,17,19,21,23},
xticklabels={2\!\!\!,4\!\!\!,6\!\!\!,8\!\!\!,10\!\!\!,12\!\!\!,14\!\!\!,16\!\!\!,18\!\!\!,20\!\!\!,22\!\!\!,24\!\!\!},
ytick={30,40,50,60},
yticklabels={30\!\!\!,40\!\!\!,50\!\!\!,60\!\!\!},
y grid style={white!69.0196078431373!black},
ylabel={\scriptsize Prices (EUR)},
ymajorgrids,
ymin=22.2051666666667, ymax=66.8248333333333,
ytick style={color=black},
xticklabel style={font=\scriptsize,rotate=90.0},
yticklabel style={font=\scriptsize},
every axis x label/.style={at={(current axis.south)}, above=-6mm, anchor=center},
every axis y label/.style={at={(current axis.west)}, left=0mm,  above=0mm, below=0mm, right=-5mm, anchor=center, rotate=90},
title style={yshift=-1.5ex},
width=\columnwidth ,height=0.5\columnwidth
]
\path [draw=color0, fill=color0]
(axis cs:0,41.3666666666667)
--(axis cs:0,31)
--(axis cs:1,31.505)
--(axis cs:2,30.9333333333333)
--(axis cs:3,29.4625)
--(axis cs:4,29.7925)
--(axis cs:5,34.5666666666667)
--(axis cs:6,31.625)
--(axis cs:7,36.8766666666667)
--(axis cs:8,34.615)
--(axis cs:9,35.6)
--(axis cs:10,35.14)
--(axis cs:11,40.6966666666667)
--(axis cs:12,37.05)
--(axis cs:13,35.5925)
--(axis cs:14,39.1233333333333)
--(axis cs:15,40.045)
--(axis cs:16,42.215)
--(axis cs:17,55.1266666666667)
--(axis cs:18,56.2166666666667)
--(axis cs:19,47.3766666666667)
--(axis cs:20,37.0666666666667)
--(axis cs:21,32.5366666666667)
--(axis cs:22,31.81)
--(axis cs:23,24.2333333333333)
--(axis cs:23,31.5366666666667)
--(axis cs:23,31.5366666666667)
--(axis cs:22,38.38)
--(axis cs:21,37.99)
--(axis cs:20,41.1)
--(axis cs:19,57.6933333333333)
--(axis cs:18,64.7966666666667)
--(axis cs:17,63.0833333333333)
--(axis cs:16,50.1725)
--(axis cs:15,48.4025)
--(axis cs:14,57.4)
--(axis cs:13,41.24)
--(axis cs:12,41.1625)
--(axis cs:11,44.6)
--(axis cs:10,44.0366666666667)
--(axis cs:9,42.69)
--(axis cs:8,38.095)
--(axis cs:7,38.99)
--(axis cs:6,33.6)
--(axis cs:5,35.4866666666667)
--(axis cs:4,31.0175)
--(axis cs:3,31.25)
--(axis cs:2,31.8333333333333)
--(axis cs:1,32.835)
--(axis cs:0,41.3666666666667)
--cycle;

\addplot [semithick, color1]
table {%
0 41.3666666666667
1 32.835
2 31.8333333333333
3 31.25
4 31.0175
5 35.4866666666667
6 33.6
7 38.99
8 38.095
9 42.69
10 44.0366666666667
11 44.6
12 41.1625
13 41.24
14 57.4
15 48.4025
16 50.1725
17 63.0833333333333
18 64.7966666666667
19 57.6933333333333
20 41.1
21 37.99
22 38.38
23 31.5366666666667
};
\addlegendentry{\scriptsize Maximum ID}
\addplot [semithick, color2]
table {%
0 31
1 31.505
2 30.9333333333333
3 29.4625
4 29.7925
5 34.5666666666667
6 31.625
7 36.8766666666667
8 34.615
9 35.6
10 35.14
11 40.6966666666667
12 37.05
13 35.5925
14 39.1233333333333
15 40.045
16 42.215
17 55.1266666666667
18 56.2166666666667
19 47.3766666666667
20 37.0666666666667
21 32.5366666666667
22 31.81
23 24.2333333333333
};
\addlegendentry{\scriptsize Minimum ID}
\addplot [semithick, color3]
table {%
0 34.5966666666667
1 32.3775
2 31.3833333333333
3 30.1125
4 30.5325
5 34.97
6 31.8975
7 37.9333333333333
8 37.27
9 39.1233333333333
10 42.2133333333333
11 42.8
12 39.6025
13 37.71
14 43.67
15 45.4775
16 48.41
17 60.3466666666667
18 61.2033333333333
19 52.6866666666667
20 39.2933333333333
21 35.9266666666667
22 35.39
23 29.6466666666667
};
\addlegendentry{\scriptsize Average ID}
\addplot [semithick, color4]
table {%
0 41.96
1 30.2775
2 27.9225
3 27.925
4 30.4625
5 34.25
6 31.1325
7 33.735
8 35.715
9 37.695
10 39.35
11 40.77
12 38.88
13 36.005
14 34.955
15 46.625
16 50.785
17 50.845
18 51.68
19 45.865
20 36.245
21 35.44
22 35
23 31.77
};
\addlegendentry{\scriptsize DA price}
\end{axis}

\end{tikzpicture}
\vspace{-2mm}
	\input{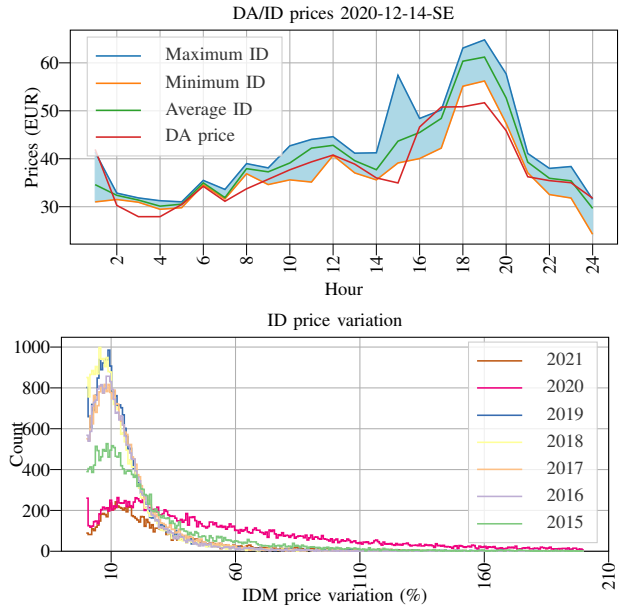}
	\caption{The DA and ID market prices in 2020-12-14 (up) and ID market price variation (down). Data source: \cite{nordpool2021data}}
	\label{fig:fig_compare_day_ahead_and_intraday_prices}
\end{figure}

To deal with the liquidity and trading risks, we need to deal with the existing and growing uncertainties in the electricity markets. Uncertainty in ID markets are caused by several factors such as changes in regulations, higher trading risk, volume-dependent prices, and etc. Compared to DA market prices, ID market prices are often more volatile. This variation is very clear by looking into historical data, e.g., it is shown for 2020-12-14 in Fig. \ref{fig:fig_compare_day_ahead_and_intraday_prices} (up). Intraday prices vary in the presented ranges for minimum and maximum prices. For instance, average ID market price at 3 pm is 43.67EUR which alters between 39.12EUR (-10.4\%) and 57.40EUR (23.9\%) in just one hour, i.e. 18.28EUR (41.8\%) variation in one hour. In addition, these changes are not similar in different periods of time. ID market price variation is shown in Fig. \ref{fig:fig_compare_day_ahead_and_intraday_prices} (down) for 2015 to 2021 which is higher in recent years. The ID market price variation calculated using the following equation can go up to 200\%.
\vspace{-2mm}
\begin{equation*}
\text{ID market~price~variation}=\dfrac{\text{Max.~price} \m \text{Min.~price}}{\text{Avg.~price}}\times100    
\end{equation*}
\vspace{-5mm}
\vspace{-5mm}

Sweden is divided into four bidding zones i.e. SE1-SE4. Sometimes, prices in these biding zones are the same. To study this, percentage of time steps when the prices are the same in entire Sweden is shown in Table \ref{tab:similar_prices} for 2015 to 2021. For instance, in 2019, DA market prices in 78.77\% of the time (6826 data points) prices in four bidding zones of Sweden are the same, while this is 5.3\% (459 data points) in ID average prices. Therefore, ID prices change more in different bidding zones. Overtime, from 2015 to 2020, this percentage is decreased in DA market and increased in ID market. This shows the growing importance of modeling ID prices in recent years.
\begin{table}[!h]
    \centering
    \caption{Percentage of similar prices in Sweden}
    \begin{tabularx}{\columnwidth}{XXXXXc}
    \toprule \rowcolor{color111}
     DA &      High &       Low &      Last &       Avg & Year \\\rowcolor{color110}
    \midrule
 85.81 &  2.98 &  2.82 &  3.66 &  1.21 & 2015 \\\rowcolor{color112}
 92.94 &  2.18 &  2.24 &  3.36 &  0.58 & 2016 \\\rowcolor{color110}
 85.45 &  2.67 &  2.84 &  3.63 &  1.82 & 2017 \\\rowcolor{color112}
 81.99 &  1.72 &  1.62 &  3.04 &  1.14 & 2018 \\\rowcolor{color110}
 78.77 &  5.69 &  5.70 &  5.84 &  5.3 & 2019 \\\rowcolor{color112}
 47.66 & 15.22 & 15.11 & 14.89 & 14.77 & 2020 \\\rowcolor{color110}
 60.26 &  0.32 &  0.16 &  0.16 &  0.00 & 2021 \\
    \bottomrule
    \end{tabularx}
    \label{tab:similar_prices}
\end{table}

\begin{figure}[!h]
	\centering
\begin{tikzpicture}

\definecolor{color0}{rgb}{0.67843137254902,0.847058823529412,0.901960784313726}
\definecolor{color1}{rgb}{0.12156862745098,0.466666666666667,0.705882352941177}
\definecolor{color2}{rgb}{1,0.498039215686275,0.0549019607843137}
\definecolor{color3}{rgb}{0.172549019607843,0.627450980392157,0.172549019607843}
\definecolor{color4}{rgb}{0.83921568627451,0.152941176470588,0.156862745098039}

\begin{axis}[
legend cell align={left},
legend style={fill opacity=0.8, draw opacity=1, text opacity=1, draw=white!80!black},
tick align=outside,
tick pos=left,
title={\scriptsize DA and ID prices in 2020-3-28-SE2},
x grid style={white!69.0196078431373!black},
xlabel={\scriptsize Hour},
xmajorgrids,
xmin=-1.15, xmax=24.15,
xtick style={color=black},
xtick={0,1,2,3,4,5,6,7,8,9,10,11,12,13,14,15,16,17,18,19,20,21,22,23},
xticklabels={1,2,3,4,5,6,7,8,9,10,11,12,13,14,15,16,17,18,19,20,21,22,23,24},
y grid style={white!69.0196078431373!black},
ylabel={\scriptsize Prices (EUR)},
ymajorgrids,
ymin=-14.188, ymax=22.948,
ytick style={color=black},
width=\columnwidth,
height=0.6\columnwidth,
xticklabel style={font=\scriptsize,rotate=90},
yticklabel style={font=\scriptsize},
title style={yshift=-1.5ex},
every axis x label/.style={at={(current axis.south)}, anchor=center, above=-8mm},
every axis y label/.style={at={(current axis.west)}, right=-7mm, anchor=center, rotate=90},
]
\path [draw=color0, fill=color0]
(axis cs:0,4.01)
--(axis cs:0,4)
--(axis cs:1,0)
--(axis cs:2,0)
--(axis cs:3,0)
--(axis cs:4,4.01)
--(axis cs:5,0)
--(axis cs:6,0)
--(axis cs:7,0)
--(axis cs:8,0)
--(axis cs:9,3)
--(axis cs:10,0)
--(axis cs:11,0)
--(axis cs:12,-12.5)
--(axis cs:13,0)
--(axis cs:14,-12.5)
--(axis cs:15,-0.45)
--(axis cs:16,0)
--(axis cs:17,0)
--(axis cs:18,0)
--(axis cs:19,0.5)
--(axis cs:20,2.5)
--(axis cs:21,1.71)
--(axis cs:22,2)
--(axis cs:23,1.9)
--(axis cs:23,8.01)
--(axis cs:23,8.01)
--(axis cs:22,6.13)
--(axis cs:21,6.9)
--(axis cs:20,6.9)
--(axis cs:19,6)
--(axis cs:18,0)
--(axis cs:17,0)
--(axis cs:16,0)
--(axis cs:15,4.55)
--(axis cs:14,4.6)
--(axis cs:13,0)
--(axis cs:12,21.26)
--(axis cs:11,0)
--(axis cs:10,0)
--(axis cs:9,5.48)
--(axis cs:8,0)
--(axis cs:7,0)
--(axis cs:6,0)
--(axis cs:5,0)
--(axis cs:4,7)
--(axis cs:3,0)
--(axis cs:2,0)
--(axis cs:1,0)
--(axis cs:0,4.01)
--cycle;

\addplot [semithick, color1]
table {%
0 4.01
1 0
2 0
3 0
4 7
5 0
6 0
7 0
8 0
9 5.48
10 0
11 0
12 21.26
13 0
14 4.6
15 4.55
16 0
17 0
18 0
19 6
20 6.9
21 6.9
22 6.13
23 8.01
};
\addlegendentry{\scriptsize Maximum ID price}
\addplot [semithick, color2]
table {%
0 4
1 0
2 0
3 0
4 4.01
5 0
6 0
7 0
8 0
9 3
10 0
11 0
12 -12.5
13 0
14 -12.5
15 -0.45
16 0
17 0
18 0
19 0.5
20 2.5
21 1.71
22 2
23 1.9
};
\addlegendentry{\scriptsize Minimum ID price}
\addplot [semithick, color3]
table {%
0 4
1 0
2 0
3 0
4 5.6
5 0
6 0
7 0
8 0
9 4.41
10 0
11 0
12 0.01
13 0
14 2.46
15 3.96
16 0
17 0
18 0
19 2.92
20 4.76
21 3.72
22 4.75
23 4.36
};
\addlegendentry{\scriptsize Average ID price}
\addplot [semithick, color4]
table {%
0 5.46
1 5.37
2 5.3
3 5.3
4 5.36
5 5.38
6 5.47
7 5.49
8 5.53
9 5.48
10 5.44
11 5.31
12 4.99
13 4.1
14 3.97
15 4.22
16 5.05
17 5.55
18 5.82
19 5.76
20 5.45
21 5.2
22 5.06
23 4.91
};
\addlegendentry{\scriptsize DA price}
\end{axis}

\end{tikzpicture}
	\caption{Compare DA and ID prices}
	\label{fig:fig_compare_day_ahead_and_intraday_prices_max}
\end{figure}
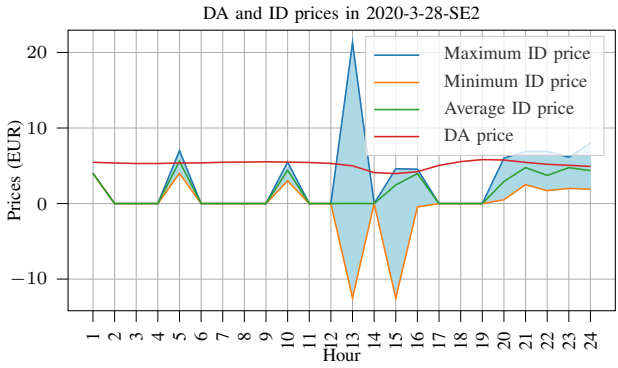

Also there is a strong correlation between prices in DA and ID markets, the ID market prices may vary significantly. For instance,  in 2020-03-28 in bidding zone SE2, DA prices are almost constant while the ID prices vary from -10 to 20 EUR/MWh as shown in Fig. \ref{fig:fig_compare_day_ahead_and_intraday_prices_max}.
\vspace{-5mm}

Also international ID markets are established from 1999 \cite{nordpool2021history}, launching several new projects lead to more changes in market regulations. For instance, European Cross-Border Intraday (XBID) and European Single Intraday Coupling (SIDC) solution (from 2018) and Local Implementation Projects are implemented in 2018 and expanded in 2019 \cite{entsoe2021sidc}. As expected, dynamics of the market changes when new regulations or new projects are implemented. Therefore, there is a continuous lack of sufficient data to implement uncertainties in ID markets. We address this by proposing three approaches to model multi-settlement prices (i.e. DA and ID prices).

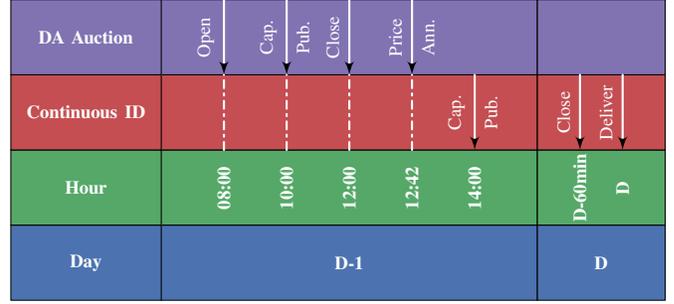
\begin{figure}[h!]
    \centering
    \begin{tikzpicture}[line join=round, line cap=round, auto]
    \pgfmathsetmacro{\tikzwd}{1.7};
    \pgfmathsetmacro{\tikzwdm}{5};
    \pgfmathsetmacro{\tikzwt}{2};
    \pgfmathsetmacro{\tikzh}{1};
    \draw[draw=black,fill=color101] ($(0,0)$) rectangle ++(\tikzwdm,\tikzh);
    \draw[draw=black,fill=color101] ($(\tikzwdm,0)$) rectangle ++(\tikzwd,\tikzh);
    \draw[draw=black,fill=color101] ($(0,0)$) rectangle ++(-\tikzwt,\tikzh);
    \draw[draw=black,fill=color102] ($(0,\tikzh)$) rectangle ++(\tikzwdm,\tikzh);
    \draw[draw=black,fill=color102] ($(\tikzwdm,\tikzh)$) rectangle ++(\tikzwd,\tikzh);
    \draw[draw=black,fill=color102] ($(0,\tikzh)$) rectangle ++(-\tikzwt,\tikzh);
    \draw[draw=black,fill=color103] ($(0,\tikzh*2)$) rectangle ++(\tikzwdm,\tikzh);
    \draw[draw=black,fill=color103] ($(\tikzwdm,\tikzh*2)$) rectangle ++(\tikzwd,\tikzh);
    \draw[draw=black,fill=color103] ($(0,\tikzh*2)$) rectangle ++(-\tikzwt,\tikzh);
    \draw[draw=black,fill=color104] ($(0,\tikzh*3)$) rectangle ++(\tikzwdm,\tikzh);
    \draw[draw=black,fill=color104] ($(\tikzwdm,\tikzh*3)$) rectangle ++(\tikzwd,\tikzh);
    \draw[draw=black,fill=color104] ($(0,\tikzh*3)$) rectangle ++(-\tikzwt,\tikzh);
    \node[text=white] at (\tikzwdm/2,\tikzh/2) {\scriptsize \textbf{D-1}};
    \node[text=white] at (\tikzwdm+\tikzwd/2,\tikzh/2) {\scriptsize \textbf{D}};
    \node[text=white] at (-\tikzwt/2,\tikzh/2) {\scriptsize \textbf{Day}};
    \node[text=white] at (-\tikzwt/2,\tikzh*3/2) {\scriptsize \textbf{Hour}};
    \node[text=white] at (-\tikzwt/2,\tikzh*5/2) {\scriptsize \textbf{Continuous ID}};
    \node[text=white] at (-\tikzwt/2,\tikzh*7/2) {\scriptsize \textbf{DA Auction}};
    \node[text=white,rotate=90] at (\tikzwdm/2,\tikzh*3/2) {\scriptsize \textbf{12:00}};
    \node[text=white,rotate=90] at (\tikzwdm/2-\tikzwdm/6,\tikzh*3/2) {\scriptsize \textbf{10:00}};
    \node[text=white,rotate=90] at (\tikzwdm/2-\tikzwdm*2/6,\tikzh*3/2) {\scriptsize \textbf{08:00}};
    \node[text=white,rotate=90] at (\tikzwdm/2+\tikzwdm/6,\tikzh*3/2) {\scriptsize \textbf{12:42}};
    \node[text=white,rotate=90] at (\tikzwdm/2+\tikzwdm*2/6,\tikzh*3/2) {\scriptsize \textbf{14:00}};
    \node[text=white,rotate=90] at (\tikzwdm+\tikzwd*1/3,\tikzh*3/2) {\scriptsize \textbf{D-60min}};
    \node[text=white,rotate=90] at (\tikzwdm+\tikzwd*2/3,\tikzh*3/2) {\scriptsize \textbf{D}};
    
    \draw [thick,dash dot,draw=white] (\tikzwdm/2,\tikzh*2) -- (\tikzwdm/2,\tikzh*3);
    \draw [thick,dash dot,draw=white] (\tikzwdm/2-\tikzwdm*1/6,\tikzh*2) -- (\tikzwdm/2-\tikzwdm*1/6,\tikzh*3);
    \draw [thick,dash dot,draw=white] (\tikzwdm/2-\tikzwdm*2/6,\tikzh*2) -- (\tikzwdm/2-\tikzwdm*2/6,\tikzh*3);
    \draw [thick,dash dot,draw=white] (\tikzwdm/2+\tikzwdm*1/6,\tikzh*2) -- (\tikzwdm/2+\tikzwdm*1/6,\tikzh*3);
    \draw [thick,-latex',draw=white] (\tikzwdm/2-\tikzwdm*2/6,\tikzh*4) -- node [rotate=90,anchor=south,text=white] {\scriptsize Open} (\tikzwdm/2-\tikzwdm*2/6,\tikzh*3);
    \draw [thick,-latex',draw=white] (\tikzwdm/2-\tikzwdm*1/6,\tikzh*4) -- node [rotate=90,anchor=south,text=white] {\scriptsize Cap.} node [rotate=90,anchor=north,text=white] {\scriptsize Pub.} (\tikzwdm/2-\tikzwdm*1/6,\tikzh*3);
    \draw [thick,-latex',draw=white] (\tikzwdm/2,\tikzh*4) -- node [rotate=90,anchor=south,text=white] {\scriptsize Close} (\tikzwdm/2,\tikzh*3);
    \draw [thick,-latex',draw=white] (\tikzwdm/2+\tikzwdm*1/6,\tikzh*4) -- node [rotate=90,anchor=south,text=white] {\scriptsize Price} node [rotate=90,anchor=north,text=white] {\scriptsize Ann.} (\tikzwdm/2+\tikzwdm*1/6,\tikzh*3);
    \draw [thick,-latex',draw=white] (\tikzwdm/2+\tikzwdm*2/6,\tikzh*3) -- node [rotate=90,anchor=south,text=white] {\scriptsize Cap.} node [rotate=90,anchor=north,text=white] {\scriptsize Pub.} (\tikzwdm/2+\tikzwdm*2/6,\tikzh*2);
    \draw [thick,-latex',draw=white] (\tikzwdm+\tikzwd*2/3,\tikzh*3) -- node [rotate=90,anchor=south,text=white] {\scriptsize Deliver}  (\tikzwdm+\tikzwd*2/3,\tikzh*2);
    \draw [thick,-latex',draw=white] (\tikzwdm+\tikzwd*1/3,\tikzh*3) -- node [rotate=90,anchor=south,text=white] {\scriptsize Close}  (\tikzwdm+\tikzwd*1/3,\tikzh*2);
    \end{tikzpicture}
    \caption{Schematic of DA and ID electricity markets}
    \label{fig:schematic}
\end{figure}

Schematic of DA and ID electricity markets, in Nord Pool electricity exchange, \cite{nordpool2021data} is shown in Fig. \ref{fig:schematic}. DA market operates as an auction which opens at 08:00 the day before the delivery. Available capacities for DA market are published at 10:00 and the auction is closed at 12:00. Then the DA market is cleared and prices are announced at 12:42. On the other hand, ID market is a continuous pay as bid market which is always open for new bids. Capacities for the new day are published at 14:00 the day before the delivery. Market participants have up to one hour before the time of delivery to use ID market for that specific time of delivery. ID markets are continuously open for bids. Therefore, market participants can use announced prices at 12:42 in D-1 (the day before delivery) and published capacities at 14:00 in D-1 to bid in the market to maximize their profits.
\vspace{-5mm}

\subsection{Background Research}\label{sec:background}
Studies on price modeling and scenario generation are reviewed in this section. Scenario-generation methods can be categorized under price modeling approaches where different scenarios are considered for uncertain parameters which is widely used in literature for stochastic programming, i.e. \cite{kaut2003evaluation} and \cite{ma2013scenario} and \cite{mitra2019regression} and \cite{wu2020spatiotemporal}.
Scenario generation algorithms can be divided into two general categories:
(1) Fitting/training a model: These algorithms aim to fit/train a model to the available historical data for the uncertain parameter. Then, these models are used to generate scenarios. These models mostly look at each single data point independently. Some of the literatures in this category are \cite{kaffash2020comparison,marcjasz2020beating,liang2019sequence,liang2019sequence,qiao2020renewable}. This category is also known as data-driven based scenario generation \cite{kaffash2020comparison} in literature.
(2) Probabilistic scenario generation: These algorithms aim to generate scenarios based on predicting probability distribution function (PDF). Later, the predicted PDF is used to generate scenarios for the uncertain parameters. Some of the literatures in this category are \cite{di2009scenario,xie2018temperature,kaffash2020comparison,ziel2018probabilistic,hu2020novel,kirui2020scentrees}. This category is also known as statistical-based scenario generation \cite{kaffash2020comparison} in literature.
Literature on price modeling are reviewed in Table \ref{tab:review}.
Recent scenario generation methods are reviewed in more details in papers such as \cite{li2020review,paul2021review}.
\vspace{-3mm}
\setlength{\tabcolsep}{1pt}
\begin{table}[!h]
	\caption{Literature on Price Modeling Approaches}\label{tab:review}
	\centering\!\!
	\begin{tabularx}{\columnwidth}{llXcX}
		\toprule
		\rowcolor{color111}
		\rotatebox[origin=c]{90}{\centering Reference} &
		{\centering Method} &
		{\centering Application} &
		\rotatebox[origin=c]{90}{\centering Model} &
		{\centering Drawbacks} \\ 
		\midrule
		\rowcolor{color110}
		\cite{kaut2003evaluation} &
		MMSG & 
		portfolio management & ST & 
		more than 1000 scenarios required   \\
		\rowcolor{color112}
		\cite{ma2013scenario} & 
		Probabilistic & Wind power & 
		ECDF, IT & 
		Employ wind power forecast  \\
		\rowcolor{color110}
		\cite{mitra2019regression} & 
		Regression & Performance management & 
		SLR & 
		limited forecasting capability  \\
		\rowcolor{color112}
		\cite{wu2020spatiotemporal} & 
		Spatiotemporal & 
		Road network traffic flow & 
		LSTM, GANs & 
		hyper parameter tuning  \\
		\rowcolor{color110}
		\cite{chen2018modelfree} & 
		Spatiotemporal & 
		Wind and solar generation & 
		GANs & 
		hyper parameter tuning \\
		\bottomrule
	\end{tabularx}\centering\\
\begin{tablenotes}
\small
\item ECDF: empirical cumulative distribution function,
GANs: generative adversarial networks,
IT: inverse transform,
LSTM: long short term memory,
MMSG: moment-matching scenario-generation,
SLR: simple linear regression,
ST: Scenario tree,
\end{tablenotes}
\end{table}
\setlength{\tabcolsep}{6pt}
\subsection{Contributions}\label{sec:contributions}
ID markets are facing structural changes and lack of insufficient data as explained in Section \ref{sec:motivation} which are addressed in this paper. In particular, we propose \emph{long short term memory (LSTM)}, \emph{deep convolutional generative adversarial networks (DCGAN)}, and \emph{No-U-Turn sampler (NUTS)} algorithms to model ID market price. The main contributions of this paper are the followings: 
(1) We have explored the DA and ID market data. We have looked into effect of time, area, and bidding zone, and volume on behavior of these markets.
(2) We have considered time series of the ID market prices. Then, we employed an LSTM-based algorithm to model the ID prices by generating different scenarios based on the latest update in the time series which is effective in capturing temporal dynamics. Advantage of this algorithm is to generate time series for prices similar to the actual data which makes it more suitable for the system operation. 
(3) In the second approach, we consider the ID prices as unknown functions (e.g. black boxes) with 24 inputs (one for each hour). Then, we look at the available information for each time step as inputs. In the DCGANs-based approach, we develop an ID price model based on this assumption. Advantage of this approach is to generate prices without fitting a PDF to the historical data, e.g. without knowledge about the probabilistic features. As a result, the prices are able to directly adapt to the changes in the market.
(4) In the third approach, e.g. NUTS-based algorithm, the ID market prices are considered as unknown random numbers. Then, we converge to a target PDF for these prices. Then, we can sample from the PDF for further market studies. Advantage of this algorithm is to generate prices with more similar PDFs by fitting a PDF to the actual data which is required in more long-term studies.
In this paper, ID prices are modeled with each approach and results are compared.

\vspace{-5mm}
\section{The proposed ID price models}\label{sec:proposedapproach}
The proposed approaches for ID price modeling are explained in this section.
\vspace{-5mm}
\subsection{Long Short Term Memory (LSTM) based model}\label{sec:lstm}
Recurrent neural networks always suffered with vanishing gradient which cause different inputs to have high/low influence on the calculated gradient when we are training these networks. LSTM networks are introduced in 1997 in \cite{hochreiter1997long} with the aim of solving this problem. Advantages of LSTM networks are long duration memory (capable of learning long-term dependencies) and a state vector which is separate from output. As discussed before, ID market prices have both short and long term dependencies. Therefore, LSTM networks are suitable for modeling ID market prices.

\begin{figure}[!h]
\centering
\begin{tikzpicture}
\pgfmathsetmacro{\tikzinnersep}{2.5};
\pgfmathsetmacro{\tikzinnersepp}{1};
\pgfmathsetmacro{\tikzinnerseps}{1};
\node[draw,circle,inner sep=\tikzinnersep pt](x) {\scriptsize $\varlstmx$};
\node[draw,rectangle,inner sep=\tikzinnersep pt, above right of=x, node distance = 10mm](w) {\scriptsize $\parlstmw$};
\node[draw,rectangle,inner sep=\tikzinnersep pt, below right of=x, node distance = 10mm](ws) {\scriptsize $\parlstmws$};
\node[draw,rectangle,inner sep=\tikzinnersep pt, above left of=x, node distance = 10mm](wi) {\scriptsize $\parlstmwi$};
\node[draw,rectangle,inner sep=\tikzinnersep pt, below left of=x, node distance = 10mm](wo) {\scriptsize $\parlstmwo$};
\node[draw,circle, inner sep=\tikzinnersepp pt, above of=w, node distance=5.5mm](wp) {\scriptsize +};
\node[draw,circle, inner sep=\tikzinnersepp pt, above of=wi, node distance=5.5mm](wip) {\scriptsize +};
\node[draw,circle, inner sep=\tikzinnersepp pt, below of=ws, node distance=5.5mm](wsp) {\scriptsize +};
\node[draw,circle, inner sep=\tikzinnersepp pt, below of=wo, node distance=5.5mm](wop) {\scriptsize +};
\node[draw, rectangle, inner sep=\tikzinnersep pt, above of=wp, node distance=5.5mm](v) {\scriptsize $\parlstmv$};
\node[draw, rectangle, inner sep=\tikzinnersep pt, above of=wip, node distance=5.5mm](vi) {\scriptsize $\parlstmvi$};
\node[draw, rectangle, inner sep=\tikzinnersep pt, below of=wsp, node distance=5.5mm](vs) {\scriptsize $\parlstmvs$};
\node[draw, rectangle, inner sep=\tikzinnersep pt, below of=wop, node distance=5.5mm](vo) {\scriptsize $\parlstmvo$};
\node[draw, rectangle, inner sep=\tikzinnersep pt, right of=wp, node distance=8.5mm](wpf) {\scriptsize $\funlstmtanh$};
\node[draw, rectangle, inner sep=\tikzinnersep pt, left of=wip, node distance=8.5mm](wipf) {\scriptsize $\funlstmsigmoid$};
\node[draw, rectangle, inner sep=\tikzinnersep pt, right of=wsp, node distance=8.5mm](wspf) {\scriptsize $\funlstmsigmoid$};
\node[draw, rectangle, inner sep=\tikzinnersep pt, left of=wop, node distance=8.5mm](wopf) {\scriptsize $\funlstmsigmoid$};
\node[draw, circle, inner sep=\tikzinnerseps pt, right of=wpf, node distance=7mm](s11) {\scriptsize };
\node[draw, circle, inner sep=\tikzinnerseps pt, right of=s11, node distance=2mm](s12) {\scriptsize };
\node[draw, circle, inner sep=0 pt, above of=x, node distance=18mm](ymu) {\scriptsize $\varlstmyminus$};
\node[draw, circle, inner sep=0 pt, below of=x, node distance=18mm](yml) {\scriptsize $\varlstmyminus$};
\node[draw,circle, inner sep=\tikzinnersepp pt, right of=s12, node distance=4mm](s1p) {\scriptsize +};
\node[draw, circle, inner sep=\tikzinnerseps pt, below of=s1p, node distance=24mm](s22) {\scriptsize };
\node[draw, circle, inner sep=\tikzinnerseps pt, below of=s22, node distance=2mm](s21) {\scriptsize };
\node[draw, circle, inner sep=0 pt, below of=s21, node distance=5mm](sm) {\scriptsize $\varlstmsminus$};
\node[draw, circle, inner sep=0 pt, right of=s1p, node distance=5mm](s) {\scriptsize $\varlstms$};
\node[draw, rectangle, inner sep=\tikzinnersep pt, right of=s, node distance=6.5mm](sf) {\scriptsize $\funlstmtanh$};
\node[draw, circle, inner sep=\tikzinnerseps pt, right of=sf, node distance=5.5mm](s31) {\scriptsize };
\node[draw, circle, inner sep=\tikzinnerseps pt, right of=s31, node distance=2mm](s32) {\scriptsize };
\node[draw, circle, inner sep=\tikzinnersep pt, right of=s32, node distance=5mm](y) {\scriptsize $\varlstmy$};
\path [draw, -latex'] (x) -- (w);
\path [draw, -latex'] (x) -- (ws);
\path [draw, -latex'] (x) -- (wi);
\path [draw, -latex'] (x) -- (wo);
\path [draw, -latex'] (w) -- (wp);
\path [draw, -latex'] (wi) -- (wip);
\path [draw, -latex'] (ws) -- (wsp);
\path [draw, -latex'] (wo) -- (wop);
\path [draw, -latex'] (v) -- (wp);
\path [draw, -latex'] (vi) -- (wip);
\path [draw, -latex'] (vs) -- (wsp);
\path [draw, -latex'] (vo) -- (wop);
\path [draw, -latex'] (wp) -- (wpf);
\path [draw, -latex'] (wip) -- (wipf);
\path [draw, -latex'] (wsp) -- (wspf);
\path [draw, -latex'] (wop) -- (wopf);
\path [draw] (wpf) -- (s11) --++ (0.2,0.2);
\path [draw, latex'-] ($(s11.north)+(0.1,0.05)$) --++ (0,0.8) -| (wipf);
\path [draw, -latex'] (ymu) -- (v);
\path [draw, -latex'] (ymu) -- (vi);
\path [draw, -latex'] (yml) -- (vs);
\path [draw, -latex'] (yml) -- (vo);
\path [draw, -latex'] (s12) -- (s1p);
\path [draw, -latex'] (s22) -- (s1p);
\path [draw] (sm) -- (s21) --++ (-0.2,0.2);
\path[draw, -latex'] (wspf) --++ (1.22,0);
\path [draw, -latex'] (s1p) -- (s);
\path [draw, -latex'] (s) -- (sf);
\path [draw] (sf) -- (s31) --++ (0.2,-0.2);
\path[draw, latex'-] ($(s31.south)+(0.1,-0.05)$) --++ (0,-3.4) -| (wopf);
\path [draw, -latex'] (s32) -- (y);
\path [] ($(s11.south)+(0.1,-0.15)$) node {\scriptsize SW1};
\path[] ($(s21.east)+(0.3,0.1)$) node {\scriptsize SW2};
\path[] ($(s31.north)+(0.1,0.15)$) node {\scriptsize SW3};
\end{tikzpicture}
\caption{Schematic of an LSTM network cell}
\label{fig:LSTM_schematic}
\end{figure}
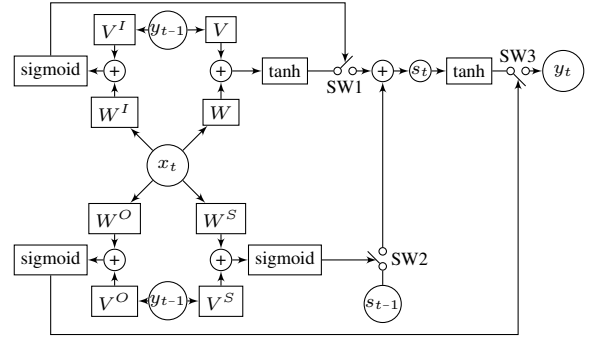

Schematic of an LSTM network cell is shown in Fig. \ref{fig:LSTM_schematic}. Initial idea is to determine results from both the current input $\varlstmx$ and previous output $\varlstmyminus$ using hyperbolic tangent as activation function $\funlstmtanh(\parlstmw\varlstmx+\parlstmv\varlstmyminus)$. Then the current output is established by three switches SW1 to SW3. 
$\parlstmwi$ and $\parlstmvi$ determine whether or not we consider the effect of the input and the previous output of the network $\varlstmyminus$ which applied by SW1.
Similarly, $\parlstmws$ and $\parlstmvs$ determine whether to skip the previous state of the network $\varlstmsminus$ or keep looking at the previous state which is applied by SW2. The ability to skip the previous state is the main advantage of LSTM networks.
Finally, $\parlstmwo$ and $\parlstmvo$ decide which parts of the current state are going to the output $\varlstmy$.

This LSTM network cell is formulated in \eqref{eq:lstm}. $\funlstmprod$ is Hadamard product or element-wise product of two matrices. Output $\varlstmy$ and current stat $\varlstms$ are calculated in \eqref{eq:lstm:y} and \eqref{eq:lstm:s}, respectively. The weight parameters $\parlstmw$, $\parlstmwi$, $\parlstmwo$, $\parlstmws$, $\parlstmv$, $\parlstmvi$, $\parlstmvo$, and $\parlstmvs$ are tuned during training of this network.
We are looking to minimize $\text{MSE}=\funlstmexpec{(\varlstmy-\varlstmyhat)^\intercal(\varlstmy-\varlstmyhat)}$ when the constraints in \eqref{eq:lstm} hold. It is a nonlinear problem but it is possible to find local optimal values for the weights using the back propagation algorithm \cite{rumelhart1986learning}.
\begin{subequations}\label{eq:lstm}
\begin{align}
&\varlstmy=\funlstmsigmoid(\parlstmwo\varlstmyminus+\parlstmvo\varlstmx+\varlstmbo)\funlstmprod\funlstmtanh(\varlstms)\label{eq:lstm:y}\\
&\varlstms=\funlstmsigmoid(\parlstmws\varlstmyminus+\parlstmvs\varlstmx+\varlstmbs)\funlstmprod\varlstmsminus+\nonumber\\
&\funlstmsigmoid(\parlstmwi\varlstmyminus\p\parlstmvi\varlstmx\p\varlstmbi)\funlstmprod\funlstmtanh(\parlstmw\varlstmyminus\p\parlstmv\varlstmx\p\varlstmb)\label{eq:lstm:s}
\end{align}
\end{subequations}

\vspace{-5mm}
\subsection{Deep Convolutional Generative Adversarial Networks (DCGANs) based model}\label{sec:dcgans}
Generative adversarial networks (GANs) are a kind of generative models based on game theory which introduced in \cite{goodfellow2014generative} as a two-model system (generator and discriminator). Each model performs a contradictory but accompanying tasks. In \cite{goodfellow2014generative}, simple multi-layer perceptrons are employed to model the generator and discriminator. Later deep convolutional generative adversarial networks (DCGANs) are introduced in \cite{radford2015unsupervised} where CNNs are employed for both the generator and discriminator models.

Generative modeling problems aim to learn the PDF which generates a training data set. Later, the estimated PDFs are employed to generate more data \cite{goodfellow2020generative}. GANs are one of the most outstanding generative models which is based on game theory. This game is between two models (generator $\labdcgansG$ and discriminator $\labdcgansD$) which play a mini-max game where players minimize their loss for the worst cases. 

\begin{figure}[!h]
\centering
\begin{tikzpicture}
\node[draw, circle](l) {\scriptsize $\vardcganslvec$};
\node[draw, trapezium, right of=l, rotate=90, minimum width=15mm, node distance=15mm](g) {\rotatebox{-90}{\scriptsize $\labdcgansG$}};
\node[right of=g, node distance=15mm](m) {};
\node[draw, circle, above of=m, node distance=5mm](xhat) {\scriptsize $\vardcgansxhatvec$};
\node[draw, trapezium, right of=m, rotate=-90, minimum width=15mm, node distance=15mm](d) {\rotatebox[]{90}{\scriptsize $\labdcgansD$}};
\node[draw, circle, below of=m, node distance=5mm](x) {\scriptsize $\vardcgansxvec$};
\node[draw, circle, right of=d, node distance=15mm](y) {\scriptsize $\vardcgansyvec$};
\path[draw, -latex'] (l)--(g);
\path[draw, -latex'] (g)--(xhat);
\path[draw, -latex'] (xhat)--(d);
\path[draw, -latex'] (x)--(d);
\path[draw, -latex'] (d)--(y);
\end{tikzpicture}
\caption{Schematic of the DCGANs-based model}
\label{fig:DCGANS_schematic}
\end{figure}
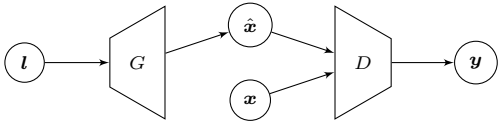

Schematic of the DCGANs network is shown in Fig \ref{fig:DCGANS_schematic}. The generator network $\labdcgansG$ provides $\vardcgansxhatvec$ for a given latent variable vector $\vardcganslvec$. Objective of the generator is to produce $\vardcgansxhatvec$ very much alike $\vardcgansxvec$. To reach this objective, parameters in the network model $\fundcgansnetG$ are adjusted during training when a noise with normal distribution is given as the latent $\vardcganslvec$.
The discriminator network $\labdcgansD$ provides an output $\vardcgansyvec$ which is one for real data ($\fundcgansnetD(\vardcgansxvec)\!=\!1$) and zero for generated data ($\fundcgansnetD\bigl(\fundcgansnetG(\vardcganslvec)\bigr)\!=\!0$). Objective of the discriminator is to distinguish real inputs $\vardcgansxvec$ from generated inputs $\vardcgansxhatvec$. To reach this objective, parameters in the network model $\fundcgansnetD$ are adjusted during training for given output of the generator network.

Training problem of the DCGANs-based model is formulated in \eqref{eq:dcgans:minmax}. In the global optimal point, $\fundcgansnetD(\vardcgansxvec)\!=\!1$ and $\fundcgansnetD\bigl(\fundcgansnetG(\vardcganslvec)\bigr)\!=\!0$ which leads to $\log\fundcgansnetD(\vardcgansxvec)\!=\!0$ and $\log\left(1-\fundcgansnetD\left(\fundcgansnetG(\vardcganslvec)\right)\right)\!=\!0$. The objective is always negative except in the global optimal point where it is zero. The discriminator tries to maximize the objective by detecting all generated data from real data. While the generator tries to minimize the objective by generating $\vardcgansxhatvec$ in such a way that it is not possible to identify by the discriminator network.
\begin{equation}\label{eq:dcgans:minmax}
\begin{aligned}
&\!\!\!\underset{\fundcgansnetG}{\text{Minimize}}\;\underset{\fundcgansnetD}{\text{Maximize}}\,\fundcgansexpec_{\vardcgansxvec}\!\!\left[\log\fundcgansnetD\!\!(\!\vardcgansxvec\!)\right]\!\!\p
\fundcgansexpec_{\vardcganslvec}\!\!\left[\log\!\!\left(\!1\m\fundcgansnetD\!\!\left(\fundcgansnetG\!(\vardcganslvec)\right)\right)\right]\!\!\!\!\!\!\!
\end{aligned}
\end{equation}
\vspace{-3mm}
\subsection{No-U-Turn Sampler (NUTS) -based model}\label{sec:nuts}
In this approach, we fit a distribution function to the real data. Therefore, we need to know overall structure of PDF of the real data from historical data which is available in ID price modeling. Then, we need to find parameters of the target PDF which will be used to take samples from. When we can not have direct samples or we do not have enough direct samples (i.e. ID market prices), this approach is especially useful. Markov Chain Monte Carlo (MCMC) method can be employed to obtain a sequence of samples from the target PDF \cite{neal2011mcmc}. Hamiltonian Monte Carlo (HMC) algorithm is an MCMC method with the advantage to avoid the random-walk behavior of MCMC and to rely less on the correlated variables \cite{nishio2019performance}. HMC is responsive to proper tuning of step-size and number of steps \cite{andrieu2008tutorial}. Recently, NUTS algorithm is introduced which improves the HMC algorithm by finding the step-size and number of steps internally \cite{hoffman2014no}.

\section{Exploring the markets}\label{sec:exploringmarkets}
DA and ID prices are explored in this section. We look at the effect of time in long-term and short-term periods. Then, we investigate the effect of area, bidding zones, and volume.

\subsection{Long-time period}\label{sec:longterm}
To study effect of long-time periods on the markets, we look at ID and DA prices in different years in Fig. \ref{fig:fig_price_avg_yearly}. Mostly distribution of these markets are similar. But, difference between these distributions is higher in 2015, 2020, and 2021 compared to other years.
\begin{figure}[h!]
	\centering
	\include{fig_day_ahead_price_avg}
	\vspace{-7mm}
	\include{fig_intraday_price_avg}
	\vspace{-7mm}
	 \caption{ID and DA market average prices from 2015 to 2021 in Sweden}
	\label{fig:fig_price_avg_yearly}
\end{figure}

\subsection{Short-time period}\label{sec:short-term}
\begin{figure}[h!]
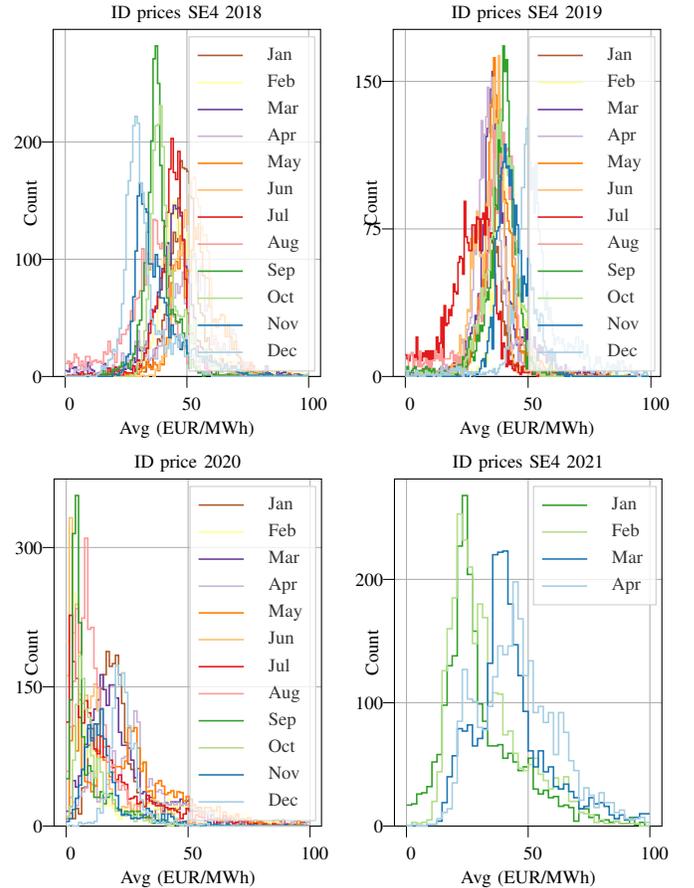

	\centering
	\input{fig_intraday_avg_2018_month}
	\input{fig_intraday_avg_2019_month}
	\input{fig_intraday_avg_2020_month}
\begin{tikzpicture}

\definecolor{color0}{rgb}{0.2,0.627450980392157,0.172549019607843}
\definecolor{color1}{rgb}{0.698039215686274,0.874509803921569,0.541176470588235}
\definecolor{color2}{rgb}{0.12156862745098,0.470588235294118,0.705882352941177}
\definecolor{color3}{rgb}{0.650980392156863,0.807843137254902,0.890196078431372}

\begin{axis}[
legend cell align={left},
legend style={fill opacity=0.8, draw opacity=1, text opacity=1, draw=white!80!black},
tick align=outside,
tick pos=left,
title={\scriptsize ID prices SE4 2021},
x grid style={white!69.0196078431373!black},
xlabel={\scriptsize Avg (EUR/MWh)},
xmajorgrids,
xmin=-4.9345, xmax=104.9445,
xtick style={color=black},
y grid style={white!69.0196078431373!black},
ylabel={\scriptsize Count},
ymajorgrids,
ymin=0, ymax=281.4,
ytick style={color=black},
width=0.58\columnwidth,
height=0.7\columnwidth,
title style={yshift=-1.5ex},
every axis x label/.style={at={(current axis.south)}, above=-7mm, anchor=center},
every axis y label/.style={at={(current axis.west)}, right=-3mm, anchor=center, rotate=90},
xticklabel style={font=\scriptsize},
yticklabel style={font=\scriptsize},
xtick={0,50,100},
xticklabels={0\!\!\!,50\!\!\!,100\!\!\!},
ytick={0,100,200,300},
yticklabels={0\!\!\!,100\!\!\!,200\!\!\!,300\!\!\!},
]
\addplot [semithick, color0, const plot mark left]
table {%
0.06 17
2.14104166666667 18
4.22208333333333 26
6.303125 27
8.38416666666667 32
10.4652083333333 52
12.54625 58
14.6272916666667 98
16.7083333333333 130
18.789375 138
20.8704166666667 227
22.9514583333333 268
25.0325 204
27.1135416666667 158
29.1945833333333 99
31.275625 85
33.3566666666667 65
35.4377083333333 65
37.51875 66
39.5997916666667 55
41.6808333333333 59
43.761875 52
45.8429166666667 45
47.9239583333333 48
50.005 55
52.0860416666667 30
54.1670833333333 26
56.248125 29
58.3291666666667 23
60.4102083333333 24
62.49125 16
64.5722916666667 17
66.6533333333333 15
68.734375 4
70.8154166666667 14
72.8964583333333 14
74.9775 7
77.0585416666667 9
79.1395833333333 4
81.220625 5
83.3016666666667 3
85.3827083333333 4
87.46375 2
89.5447916666667 2
91.6258333333333 7
93.706875 1
95.7879166666667 3
97.8689583333333 0
99.95 0
};
\addlegendentry{\scriptsize Jan}
\addplot [semithick, color1, const plot mark left]
table {%
0.06 0
2.14104166666667 3
4.22208333333333 2
6.303125 3
8.38416666666667 5
10.4652083333333 20
12.54625 44
14.6272916666667 126
16.7083333333333 160
18.789375 185
20.8704166666667 253
22.9514583333333 232
25.0325 210
27.1135416666667 148
29.1945833333333 180
31.275625 169
33.3566666666667 101
35.4377083333333 108
37.51875 107
39.5997916666667 74
41.6808333333333 77
43.761875 59
45.8429166666667 47
47.9239583333333 45
50.005 50
52.0860416666667 55
54.1670833333333 42
56.248125 37
58.3291666666667 30
60.4102083333333 39
62.49125 30
64.5722916666667 40
66.6533333333333 20
68.734375 20
70.8154166666667 11
72.8964583333333 10
74.9775 9
77.0585416666667 3
79.1395833333333 5
81.220625 2
83.3016666666667 2
85.3827083333333 4
87.46375 2
89.5447916666667 2
91.6258333333333 2
93.706875 2
95.7879166666667 0
97.8689583333333 0
99.95 0
};
\addlegendentry{\scriptsize Feb}
\addplot [semithick, color2, const plot mark left]
table {%
0.06 0
2.14104166666667 0
4.22208333333333 0
6.303125 0
8.38416666666667 1
10.4652083333333 1
12.54625 4
14.6272916666667 19
16.7083333333333 40
18.789375 57
20.8704166666667 79
22.9514583333333 77
25.0325 82
27.1135416666667 65
29.1945833333333 71
31.275625 79
33.3566666666667 132
35.4377083333333 220
37.51875 222
39.5997916666667 223
41.6808333333333 180
43.761875 147
45.8429166666667 122
47.9239583333333 93
50.005 61
52.0860416666667 48
54.1670833333333 61
56.248125 44
58.3291666666667 48
60.4102083333333 32
62.49125 30
64.5722916666667 34
66.6533333333333 28
68.734375 23
70.8154166666667 14
72.8964583333333 16
74.9775 23
77.0585416666667 18
79.1395833333333 19
81.220625 16
83.3016666666667 9
85.3827083333333 6
87.46375 9
89.5447916666667 5
91.6258333333333 5
93.706875 7
95.7879166666667 10
97.8689583333333 10
99.95 10
};
\addlegendentry{\scriptsize Mar}
\addplot [semithick, color3, const plot mark left]
table {%
0.06 0
2.14104166666667 1
4.22208333333333 0
6.303125 0
8.38416666666667 1
10.4652083333333 1
12.54625 6
14.6272916666667 11
16.7083333333333 14
18.789375 35
20.8704166666667 85
22.9514583333333 127
25.0325 103
27.1135416666667 101
29.1945833333333 80
31.275625 78
33.3566666666667 97
35.4377083333333 121
37.51875 146
39.5997916666667 139
41.6808333333333 146
43.761875 198
45.8429166666667 168
47.9239583333333 122
50.005 134
52.0860416666667 92
54.1670833333333 97
56.248125 84
58.3291666666667 92
60.4102083333333 58
62.49125 92
64.5722916666667 76
66.6533333333333 53
68.734375 45
70.8154166666667 21
72.8964583333333 31
74.9775 25
77.0585416666667 15
79.1395833333333 27
81.220625 19
83.3016666666667 19
85.3827083333333 11
87.46375 13
89.5447916666667 6
91.6258333333333 5
93.706875 6
95.7879166666667 7
97.8689583333333 3
99.95 3
};
\addlegendentry{\scriptsize Apr}
\end{axis}

\end{tikzpicture}\vspace{-3mm}
	\caption{ID average prices in different months from 2018 to 2021}
	\label{fig:fig_intraday_avg_2018_2021_month}
\end{figure}

To study effect of short-time periods on ID prices, we look at the average ID prices in different months from 2018 to 2021 in Fig. \ref{fig:fig_intraday_avg_2018_2021_month}. Distribution of the prices are significantly changed at each year. For instance, prices in July were about 45, 30, and 15 EUR/MWh in 2018, 2019, and 2020 respectively.

\subsection{Area}\label{sec:area}
\begin{figure}[!h]
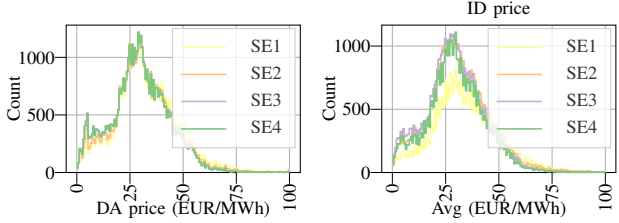

	\centering
	\input{fig_day_ahead_price_se}\vspace{-3mm}
	\input{fig_intraday_price_seAvg}
	 \caption{DA (left) and ID (right) price (EUR/MWh) in four bidding zones of Sweden (EUR/MWh)}
	\label{fig:fig_day_ahead_price_se:fig_intraday_price_avg_se}
\end{figure}

Sweden is divided into four area SE1 to SE4. Here effect of location is studied for both DA and ID electricity markets. Average DA market prices (EUR/MWh) are shown in Fig. \ref{fig:fig_day_ahead_price_se:fig_intraday_price_avg_se} (left) in SE1 to SE4 in Sweden. Distribution of the prices are mainly similar in Sweden DA market.
Average ID market prices (EUR/MWh) are shown in Fig. \ref{fig:fig_day_ahead_price_se:fig_intraday_price_avg_se} (right) in SE1 to SE4 in Sweden. Distribution of the prices is different in SE4 where prices are higher compared to other three zones. Therefore, prices in ID electricity market depend more on the location compared to the DA electricity markets.

\subsection{Bidding zones}\label{sec:biddingzone}
\begin{figure}[!h]
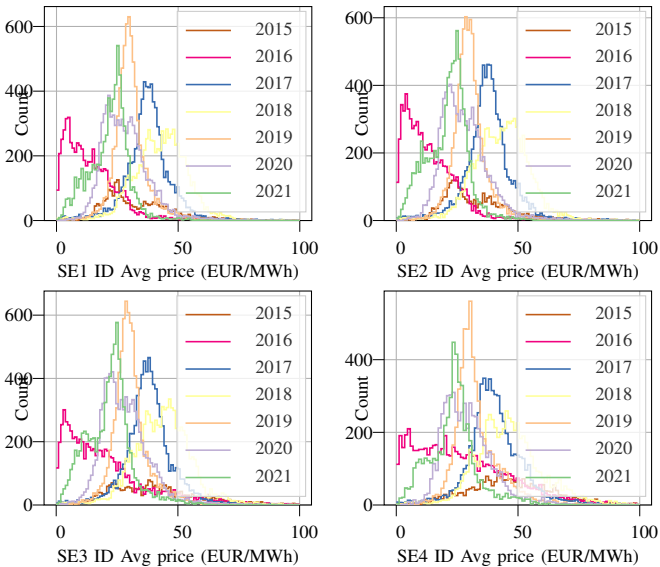

	\centering
	\input{fig_intraday_avg_seSE1_Avg}
   	\input{fig_intraday_avg_seSE2_Avg}
   	\input{fig_intraday_avg_seSE3_Avg}
   	\input{fig_intraday_avg_seSE4_Avg}\vspace{-3mm}
	 \caption{ID average price in SE1 to SE4 in Sweden from 2015 to 2021}
	\label{fig:fig_intraday_avg_se}
\end{figure}

Here we look at four zones of Sweden (SE1 to SE4) and variation of ID prices. In Fig. \ref{fig:fig_intraday_avg_se}, variation of ID prices is shown for SE1 to SE4 from 2015 to 2021. ID market prices were similar in different zones from 2015 to 2018. While, distribution of these prices vary in different zones in later years (2019 to 2021). 

\vspace{-5mm}
\subsection{Volume}\label{volume}
\begin{figure}[h!]
	\centering
	\input{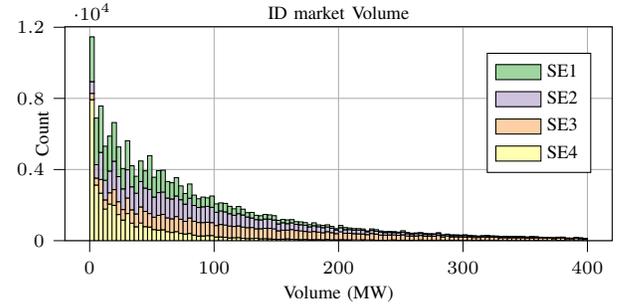}
	\caption{ID market volumes in four regions of Sweden ($0<$ volume $<400$MW)}
	\label{fig:fig_intraday_volume_se}
\end{figure}

Up to now, we have have focused on the market prices. We have looked dynamics of volume in ID market in Fig. \ref{fig:fig_intraday_volume_se} in four regions of Sweden which are between 0 and 400 (volumes=0 is removed from this figure). It is shown that transactions with smaller volume are popular in this market. Volumes less than 50MW are more common in SE4. That could be due to better connection to the rest of Europe in this region. On the other hand, volumes more than 200MW are more more common in SE3 which can be a result of connection to Norway.

Summation of volumes in DA and ID markets are compared in Table \ref{tab:volcomp} from 2021-04-24 to 2021-05-24. ID market volumes are between 1.9\% to 6.6\% of volumes in DA market in the same time period.
\begin{table}[h!]
	\caption{Buy and sell volumes in DA and ID electricity markets from 2021-04-24 to 2021-05-24}\label{tab:volcomp}
	\centering\!\!
	\begin{tabularx}{\columnwidth}{cYYYY}
		\toprule
		\rowcolor{color111}
		{\centering Zone} &
		{\centering Type} &
		{\centering DA volume (MW)} &
		{\centering ID volume (MW)} &
		{\centering ID volume (\%)} \\ 
		\midrule
		\rowcolor{color110}
    	SE1	&	Buy	&	865 237.5	&	18 832.4	&	2.2	\\ \rowcolor{color112}
    	SE1	&	Sell	&	2 086 117.5	&	40 491.2	&	1.9	\\ \rowcolor{color110}
    	SE2	&	Buy	&	915 577.1	&	60 801.5	&	6.6	\\ \rowcolor{color112}
    	SE2	&	Sell	&	3 403 113.4	&	81 473.3	&	2.4	\\ \rowcolor{color110}
    	SE3	&	Buy	&	5 638 029.1	&	72 795.2	&	1.3	\\ \rowcolor{color112}
    	SE3	&	Sell	&	4 570 174.9	&	66 343.0	&	1.5	\\ \rowcolor{color110}
    	SE4	&	Buy	&	1 578 820.0	&	35 793.1	&	2.3	\\ \rowcolor{color112}
    	SE4	&	Sell	&	280 141.5	&	11 692.8	&	4.2	\\
		\bottomrule
	\end{tabularx}\centering\\
\end{table}

\vspace{-5mm}
\section{Results}\label{sec:results}
\vspace{-3mm}
Our proposed algorithms have been applied to the recent ID market data. Here we are looking at more recent data (i.e. 2020 and 2021).

\vspace{-5mm}
\subsection{Long Short Term Memory (LSTM)}\label{sec:results:lstm}
\begin{figure}[!h]
	\centering
\begin{tikzpicture}

\definecolor{color1}{rgb}{1,0.498039215686275,0.0549019607843137}
\definecolor{color0}{rgb}{0.12156862745098,0.466666666666667,0.705882352941177}

\begin{axis}[
legend cell align={left},
legend style={fill opacity=0.8, draw opacity=1, text opacity=1, draw=white!80.0!black},
tick align=outside,
tick pos=left,
x grid style={lightgray!92.02614379084967!black},
title={\scriptsize LSTM-based model},
xlabel={\scriptsize Average},
xmajorgrids,
xmin=-12.226, 
xmax=150,
y grid style={lightgray!92.02614379084967!black},
ylabel={\scriptsize Density},
ymajorgrids,
ymin=0, 
ymax=0.03,
ytick style={color=black},
xticklabel style={font=\scriptsize},
yticklabel style={font=\scriptsize},
width=\columnwidth, height=0.4\columnwidth,
xtick={0,50,100,150},
xticklabels={0\!\!\!,50\!\!\!,100\!\!\!,150\!\!\!},
ytick={0,0.01,0.02,0.03},
yticklabels={0\!\!\!,0.01\!\!\!,0.02\!\!\!,0.03\!\!\!},
scaled y ticks = false,
legend style={/tikz/every odd column/.append style={column sep=0mm},},
title style={yshift=-1.5ex},
every axis x label/.style={at={(current axis.south)}, above=-3mm, anchor=center},
every axis y label/.style={at={(current axis.west)}, right=-8mm, anchor=center, rotate=90},
]
\addplot [semithick, color0, fill opacity=0]
table{
0 0
0 0.00515537878430796
8.15066666666667 0.00515537878430796
8.15066666666667 0.00088811147124633
16.3013333333333 0.00088811147124633
16.3013333333333 0.00989919371608713
24.452 0.00989919371608713
24.452 0.0144047348385075
32.6026666666667 0.0144047348385075
32.6026666666667 0.0259285227093136
40.7533333333333 0.0259285227093136
40.7533333333333 0.0262751027956536
48.904 0.0262751027956536
48.904 0.0143830735831113
57.0546666666667 0.0143830735831113
57.0546666666667 0.0100724837592572
65.2053333333333 0.0100724837592572
65.2053333333333 0.00571857142461051
73.356 0.00571857142461051
73.356 0.00372573592815533
81.5066666666667 0.00372573592815533
81.5066666666667 0.00212280302883269
89.6573333333333 0.00212280302883269
89.6573333333333 0.00110472402520885
97.808 0.00110472402520885
97.808 0.00101807900362384
105.958666666667 0.00101807900362384
105.958666666667 0.000628176406491306
114.109333333333 0.000628176406491306
114.109333333333 0.000476547618717543
122.26 0.000476547618717543
122.26 0.000194951298566268
130.410666666667 0.000194951298566268
130.410666666667 0.00010830627698126
138.561333333333 0.00010830627698126
138.561333333333 0.000151628787773764
146.712 0.000151628787773764
146.712 6.49837661887558e-05
154.862666666667 6.49837661887558e-05
154.862666666667 2.16612553962519e-05
163.013333333333 2.16612553962519e-05
163.013333333333 6.49837661887558e-05
171.164 6.49837661887558e-05
171.164 0.000129967532377512
179.314666666667 0.000129967532377512
179.314666666667 2.16612553962519e-05
187.465333333333 2.16612553962519e-05
187.465333333333 6.49837661887556e-05
195.616 6.49837661887556e-05
195.616 2.16612553962519e-05
203.766666666667 2.16612553962519e-05
203.766666666667 0
211.917333333333 0
211.917333333333 0
220.068 0
220.068 2.16612553962519e-05
228.218666666667 2.16612553962519e-05
228.218666666667 0
236.369333333333 0
236.369333333333 2.1661255396252e-05
244.52 2.1661255396252e-05
244.52 0
};
\addplot [semithick, color1, fill opacity=0]
table{
0.389787673950195 0
0.389787673950195 0.00129185121087909
4.48980331420898 0.00129185121087909
4.48980331420898 0.00215308535146515
8.58981895446777 0.00215308535146515
8.58981895446777 0.0026267647397798
12.6898336410522 0.0026267647397798
12.6898336410522 0.00383249103416057
16.7898502349854 0.00383249103416057
16.7898502349854 0.00934439042535875
20.8898658752441 0.00934439042535875
20.8898658752441 0.0163203945564384
24.9898796081543 0.0163203945564384
24.9898796081543 0.0171385593976626
29.0898952484131 0.0171385593976626
29.0898952484131 0.0232102600887943
33.1899108886719 0.0232102600887943
33.1899108886719 0.0273872329299533
37.2899284362793 0.0273872329299533
37.2899284362793 0.0256647893288583
41.3899421691895 0.0256647893288583
41.3899421691895 0.019980641356695
45.4899559020996 0.019980641356695
45.4899559020996 0.0190332656525776
49.589973449707 0.0190332656525776
49.589973449707 0.012875456391491
53.6899871826172 0.012875456391491
53.6899871826172 0.0105931750012083
57.7900047302246 0.0105931750012083
57.7900047302246 0.0117127897608212
61.8900184631348 0.0117127897608212
61.8900184631348 0.00835396727737569
65.9900360107422 0.00835396727737569
65.9900360107422 0.00693293805695668
70.0900497436523 0.00693293805695668
70.0900497436523 0.00469372824974086
74.1900634765625 0.00469372824974086
74.1900634765625 0.00383248657692981
78.2900848388672 0.00383248657692981
78.2900848388672 0.003832493708504
82.3900985717773 0.003832493708504
82.3900985717773 0.00322962952963821
86.4901123046875 0.00322962952963821
86.4901123046875 0.00155022217422634
90.5901260375977 0.00155022217422634
90.5901260375977 0.0015071604471645
94.6901397705078 0.0015071604471645
94.6901397705078 0.00167940423034003
98.7901611328125 0.00167940423034003
98.7901611328125 0.00133491353891713
102.890174865723 0.00133491353891713
102.890174865723 0.00107654317654607
106.990188598633 0.00107654317654607
106.990188598633 0.000818172814175012
111.090202331543 0.000818172814175012
111.090202331543 0.00068898635090873
115.190223693848 0.00068898635090873
115.190223693848 0.00047367899768027
119.290237426758 0.00047367899768027
119.290237426758 0.000732049360051327
123.390251159668 0.000732049360051327
123.390251159668 0
};
\addlegendentry{\scriptsize Actual\!\!\!}
\addlegendentry{\scriptsize Generated\!\!}
\end{axis}

\end{tikzpicture}\vspace{-3mm}
	 \caption{PDF of average ID prices (EUR/MWh) in LSTM-based approach}
	\label{fig:fig_lstm_pdf_Avg}
\end{figure}
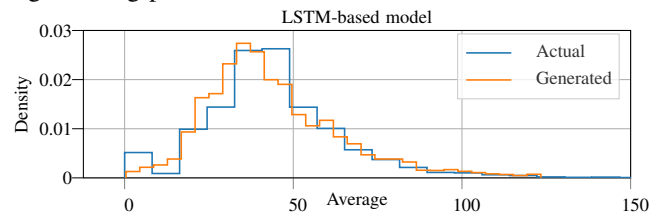

PDF of the ID average prices are shown in Fig. \ref{fig:fig_lstm_pdf_Avg} for the LSTM-based approach. Here the market data for 2020 and 2021 are employed for training and testing of the LSTM network respectively. In this approach, we look at the previous ID average prices as input and develop a model to generate prices for the next time step. The LSTM-based model is trained first with the training data (2020). Then, we generate prices for the test data (2021). PDF of the actual ID average prices and PDF of the generated ID average prices are shown in Fig. \ref{fig:fig_lstm_pdf_Avg} which shows proper performance of the LSTM-based model in generating prices with similar PDFs.
\vspace{-5mm}

\begin{figure}[!h]
	\centering
\begin{tikzpicture}

\definecolor{color1}{rgb}{1,0.498039215686275,0.0549019607843137}
\definecolor{color0}{rgb}{0.12156862745098,0.466666666666667,0.705882352941177}

\begin{axis}[
legend cell align={left},
legend style={fill opacity=0.8, draw opacity=1, text opacity=1, at={(0.03,0.97)}, anchor=north west, draw=white!80.0!black},
tick align=outside,
tick pos=left,
x grid style={lightgray!92.02614379084967!black},
xlabel={\scriptsize Hour},
xmajorgrids,
xmin=-1.15, xmax=24.15,
xtick style={color=black},
xtick={0,1,2,3,4,5,6,7,8,9,10,11,12,13,14,15,16,17,18,19,20,21,22,23},
xticklabel style = {rotate=90.0},
xticklabels={1\!\!\!,2\!\!\!,3\!\!\!,4\!\!\!,5\!\!\!,6\!\!\!,7\!\!\!,8\!\!\!,9\!\!\!,10\!\!\!,11\!\!\!,12\!\!\!,13\!\!\!,14\!\!\!,15\!\!\!,16\!\!\!,17\!\!\!,18\!\!\!,19\!\!\!,20\!\!\!,21\!\!\!,22\!\!\!,23\!\!\!,24\!\!\!},
y grid style={lightgray!92.02614379084967!black},
ylabel={\scriptsize Average},
ymajorgrids,
ymin=-4.80828056335449, ymax=100.973891830444,
ytick style={color=black},
width=\columnwidth,
height=0.4\columnwidth,
title style={yshift=-1.5ex},
every axis x label/.style={at={(current axis.south)}, above=-5mm, anchor=center},
every axis y label/.style={at={(current axis.west)}, right=-6mm, anchor=center, rotate=90},
xticklabel style={font=\scriptsize},
yticklabel style={font=\scriptsize},
ytick={0,25,50,75,100},
yticklabels={0\!\!\!,25\!\!\!,50\!\!\!,75\!\!\!,100\!\!\!},
]
\addplot [semithick, color0]
table {%
0 38.05
1 21.81
2 35.2
3 0
4 40.1
5 43.14
6 39.78
7 68.03
8 69.11
9 70.7
10 75.3
11 81.74
12 75.38
13 73.37
14 77.92
15 76.98
16 94.16
17 83.53
18 74.43
19 65.28
20 59.54
21 56
22 46.21
23 0
};
\addlegendentry{\scriptsize Validation}
\addplot [semithick, color1]
table {%
0 28.1591567993164
1 29.6325798034668
2 21.1929054260254
3 33.0848770141602
4 18.1788940429688
5 25.0667953491211
6 45.4540023803711
7 45.4156303405762
8 74.2505493164062
9 69.684684753418
10 67.0168685913086
11 76.7680130004883
12 84.8901748657227
13 76.1501083374023
14 74.6875381469727
15 82.3316879272461
16 79.1595764160156
17 96.1656112670898
18 77.3578033447266
19 68.2014236450195
20 59.6849746704102
21 53.2168884277344
22 50.0856399536133
23 39.6090965270996
};
\addlegendentry{\scriptsize Prediction}
\end{axis}

\end{tikzpicture}\vspace{-3mm}
	 \caption{Generated ID market average prices (EUR/MWh) in the LSTM-based approach for 2021-01-13}
	\label{fig:fig_lstm_pred_Avg}
\end{figure}
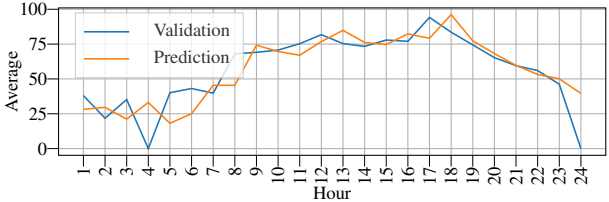

\begin{figure}[!h]
	\centering
	\input{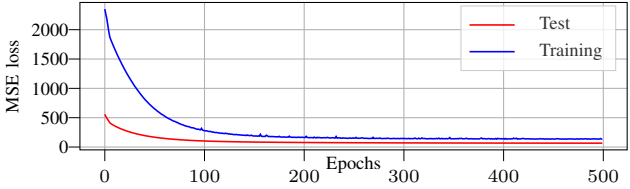}\vspace{-3mm}
	\caption{MSE loss in LSTM-based approach for training and testing data of average ID prices}
	\label{fig:fig_lstm_loss_Avg}
\end{figure}

Generated ID market average prices employing the LSTM-based approach are shown in Fig. \ref{fig:fig_lstm_pred_Avg} for 2021-01-13. Profile the generated prices follow the real prices in the 24 hour example. The main advantage of the LSTM-based approach is good performance of this approach in generating similar profile of prices. Training the proposed LSTM-based model is an iterative approach. Fig. \ref{fig:fig_lstm_loss_Avg} shows mean square error (MSE) for both training and test data for 500 iterations. Both errors are in their minimum after 200 iterations.

\subsection{Deep Convolutional Generative Adversarial Networks (DCGANs)}\label{sec:results:dcgans}
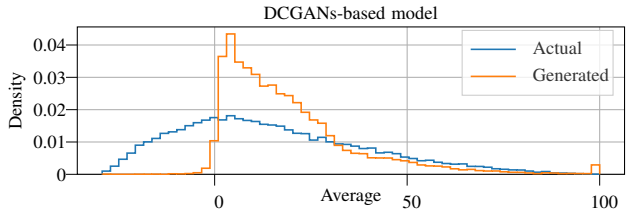
\begin{figure}[!h]
	\centering
\begin{tikzpicture}

\definecolor{color1}{rgb}{1 0.498039215686275, 0.0549019607843137}
\definecolor{color0}{rgb}{0.12156862745098, 0.466666666666667, 0.705882352941177}

\begin{axis}[
legend cell align={left} ,
legend style={fill opacity=0.8,  draw opacity=1,  text opacity=1,  draw=white!80.0!black},
tick align=outside,
tick pos=left,
title={\scriptsize DCGANs-based model},
x grid style={lightgray!92.02614379084967!black},
xlabel={\scriptsize Average},
xmajorgrids,
xmin=-35.6705003814697,
xmax=106.460508010864,
xtick style={color=black},
y grid style={lightgray!92.02614379084967!black},
ylabel={\scriptsize Density},
ymajorgrids ,
ymin=0,  ymax=0.0455910038697231,
ytick style={color=black},
xticklabel style={font=\scriptsize},
yticklabel style={font=\scriptsize},
width=\columnwidth, 
height=0.4\columnwidth,
xtick={0, 50, 100}, 
xticklabels={0\!\!\!, 50\!\!\!, 100\!\!\!} ,
ytick={0, 0.01, 0.02, 0.03, 0.04} ,
yticklabels={0\!\!\!, 0.01\!\!\!, 0.02\!\!\!, 0.03\!\!\!, 0.04\!\!\!} ,
scaled y ticks = false ,
legend style={/tikz/every odd column/.append style={column sep=-1mm},},
title style={yshift=-1.5ex},
every axis x label/.style={at={(current axis.south)}, above=-3mm, anchor=center},
every axis y label/.style={at={(current axis.west)}, right=-8mm, anchor=center, rotate=90},
]
\addplot [semithick, color0, fill opacity=0]
table{
-29.2099990844727 0
-29.2099990844727 0.000997059881581803
-27.0564994812012 0.000997059881581803
-27.0564994812012 0.00251959503672324
-24.9029979705811 0.00251959503672324
-24.9029979705811 0.00466192863550411
-22.7494983673096 0.00466192863550411
-22.7494983673096 0.00656173192338295
-20.5959987640381 0.00656173192338295
-20.5959987640381 0.00901396028078684
-18.4424991607666 0.00901396028078684
-18.4424991607666 0.0100379588361434
-16.2889976501465 0.0100379588361434
-16.2889976501465 0.0111023965192352
-14.135498046875 0.0111023965192352
-14.135498046875 0.0126653552525256
-11.9819984436035 0.0126653552525256
-11.9819984436035 0.0131369318112774
-9.82849788665771 0.0131369318112774
-9.82849788665771 0.0138106236801185
-7.67499780654907 0.0138106236801185
-7.67499780654907 0.0150367375873291
-5.52149772644043 0.0150367375873291
-5.52149772644043 0.0159664283521371
-3.36799764633179 0.0159664283521371
-3.36799764633179 0.0167613813249439
-1.21449768543243 0.0167613813249439
-1.21449768543243 0.0175428605185506
0.939002454280853 0.0175428605185506
0.939002454280853 0.0168287502209445
3.09250259399414 0.0168287502209445
3.09250259399414 0.0181222330241556
5.24600267410278 0.0181222330241556
5.24600267410278 0.0171925422593476
7.39950275421143 0.0171925422593476
7.39950275421143 0.0166670611800514
9.55300331115723 0.0166670611800514
9.55300331115723 0.0163436977886315
11.7065029144287 0.0163436977886315
11.7065029144287 0.0154678950967656
13.8600034713745 0.0154678950967656
13.8600034713745 0.0152523146772985
16.0135040283203 0.0152523146772985
16.0135040283203 0.0149020030949929
18.1670036315918 0.0149020030949929
18.1670036315918 0.0137432578272086
20.3205032348633 0.0137432578272086
20.3205032348633 0.0126923028168927
22.4740028381348 0.0126923028168927
22.4740028381348 0.0125845014133663
24.6275043487549 0.0125845014133663
24.6275043487549 0.0105903927962608
26.7810039520264 0.0105903927962608
26.7810039520264 0.0114392410738237
28.9345035552979 0.0114392410738237
28.9345035552979 0.0100514415089193
31.0880031585693 0.0100514415089193
31.0880031585693 0.00940469163440012
33.2415046691895 0.00940469163440012
33.2415046691895 0.00925648836009053
35.3950042724609 0.00925648836009053
35.3950042724609 0.00871753707274901
37.5485038757324 0.00871753707274901
37.5485038757324 0.00797647905265443
39.7020034790039 0.00797647905265443
39.7020034790039 0.00812469065667334
41.8555030822754 0.00812469065667334
41.8555030822754 0.00662908909156895
44.0090065002441 0.00662908909156895
44.0090065002441 0.00681773378487017
46.1625061035156 0.00681773378487017
46.1625061035156 0.00638657275499696
48.3160057067871 0.00638657275499696
48.3160057067871 0.005335617744681
50.4695053100586 0.005335617744681
50.4695053100586 0.00489098293262425
52.6230049133301 0.00489098293262425
52.6230049133301 0.00420382004126382
54.7765045166016 0.00420382004126382
54.7765045166016 0.00435203164528274
56.930004119873 0.00435203164528274
56.930004119873 0.00373223766483999
59.0835037231445 0.00373223766483999
59.0835037231445 0.00334149206241687
61.2370071411133 0.00334149206241687
61.2370071411133 0.00313939124876433
63.3905067443848 0.00313939124876433
63.3905067443848 0.00322023394186555
65.5440063476562 0.00322023394186555
65.5440063476562 0.00246569777186406
67.697509765625 0.00246569777186406
67.697509765625 0.00243875889522435
69.8510055541992 0.00243875889522435
69.8510055541992 0.00219622260554012
72.004508972168 0.00219622260554012
72.004508972168 0.0017246471745233
74.1580047607422 0.0017246471745233
74.1580047607422 0.00156295596467886
76.3115081787109 0.00156295596467886
76.3115081787109 0.00137432821719825
78.4650039672852 0.00137432821719825
78.4650039672852 0.00133390207330351
80.6185073852539 0.00133390207330351
80.6185073852539 0.00101053545382225
82.7720031738281 0.00101053545382225
82.7720031738281 0.000660214157493656
84.9255065917969 0.000660214157493656
84.9255065917969 0.000862320532236611
87.0790100097656 0.000862320532236611
87.0790100097656 0.000431161793630825
89.2325057983398 0.000431161793630825
89.2325057983398 0.000390738991169715
91.3860092163086 0.000390738991169715
91.3860092163086 0.000296423733121192
93.5395050048828 0.000296423733121192
93.5395050048828 0.000202106374742956
95.6930084228516 0.000202106374742956
95.6930084228516 0.000148211866560596
97.8465042114258 0.000148211866560596
97.8465042114258 4.04212749485912e-05
100.000007629395 4.04212749485912e-05
100.000007629395 0
};
\addplot [semithick, color1, fill opacity=0]
table{
-29.21 0
-29.21 5.08034364494352e-05
-27.0565 5.08034364494352e-05
-27.0565 0
-24.903 0
-24.903 0
-22.7495 0
-22.7495 0
-20.596 0
-20.596 3.38689576329568e-05
-18.4425 3.38689576329568e-05
-18.4425 1.69344788164784e-05
-16.289 1.69344788164784e-05
-16.289 5.08034364494352e-05
-14.1355 5.08034364494352e-05
-14.1355 0
-11.982 0
-11.982 0.000169344788164784
-9.8285 0.000169344788164784
-9.8285 0.000135475830531827
-7.675 0.000135475830531827
-7.675 0.000237082703430698
-5.5215 0.000237082703430698
-5.5215 0.000491099885677873
-3.368 0.000491099885677873
-3.368 0.0018797271486291
-1.2145 0.0018797271486291
-1.2145 0.0103808355145013
0.939 0.0103808355145013
0.939 0.036459932891878
3.0925 0.036459932891878
3.0925 0.0434200036854506
5.246 0.0434200036854506
5.246 0.0347156815737807
7.3995 0.0347156815737807
7.3995 0.0329375612980505
9.553 0.0329375612980505
9.553 0.0309054238400731
11.7065 0.0309054238400731
11.7065 0.0272814453733467
13.86 0.0272814453733467
13.86 0.0276032004708599
16.0135 0.0276032004708599
16.0135 0.0249106183390397
18.167 0.0249106183390397
18.167 0.0243517805380959
20.3205 0.0243517805380959
20.3205 0.0221672327707702
22.474 0.0221672327707702
22.474 0.01918676449907
24.6275 0.01918676449907
24.6275 0.016815937464763
26.781 0.016815937464763
26.781 0.0157998687357743
28.9345 0.0157998687357743
28.9345 0.0119557420444338
31.088 0.0119557420444338
31.088 0.00895833929391707
33.2415 0.00895833929391707
33.2415 0.00740036724280106
35.395 0.00740036724280106
35.395 0.00650283986552768
37.5485 0.00650283986552768
37.5485 0.00638429851381238
39.702 0.00638429851381238
39.702 0.00511421260257649
41.8555 0.00511421260257649
41.8555 0.00506340916612702
44.009 0.00506340916612702
44.009 0.00504647468731058
46.1625 0.00504647468731058
46.1625 0.0045384403228162
48.316 0.0045384403228162
48.316 0.00418281626767018
50.4695 0.00418281626767018
50.4695 0.0036747819031758
52.623 0.0036747819031758
52.623 0.00289579587761781
54.7765 0.00289579587761781
54.7765 0.00260790973773766
56.93 0.00260790973773766
56.93 0.00254017182247177
59.0835 0.00254017182247177
59.0835 0.00228615464022459
61.237 0.00228615464022459
61.237 0.00171038236046431
63.3905 0.00171038236046431
63.3905 0.00138862726295123
65.544 0.00138862726295123
65.544 0.00152410309348305
67.6975 0.00152410309348305
67.6975 0.00115154455952054
69.851 0.00115154455952054
69.851 0.0010160687289887
72.0045 0.0010160687289887
72.0045 0.00103300320780518
74.158 0.00103300320780518
74.158 0.000778986025558014
76.3115 0.000778986025558014
76.3115 0.000626575716209699
78.465 0.000626575716209699
78.465 0.00060964123739322
80.6185 0.00060964123739322
80.6185 0.000423361970411959
82.772 0.000423361970411959
82.772 0.000220148224614221
84.9255 0.000220148224614221
84.9255 0.000457230928044915
87.079 0.000457230928044915
87.079 0.000389493012779002
89.2325 0.000389493012779002
89.2325 0.0002370827034307
91.386 0.0002370827034307
91.386 0.000270951661063653
93.5395 0.000270951661063653
93.5395 0.000237082703430697
95.693 0.000237082703430697
95.693 0.000220148224614218
97.8465 0.000220148224614218
97.8465 0.00291273035643431
100 0.00291273035643431
100 0
};
\addlegendentry{\scriptsize Actual\!\!\!}
\addlegendentry{\scriptsize Generated\!\!\!}
\end{axis}

\end{tikzpicture}\vspace{-3mm}
	 \caption{PDF of average ID prices (EUR/MWh) in DCGANs-based approach}
	\label{fig:fig_pdf_dcgen_Avg}
\end{figure}

PDF of the ID average prices are shown in Fig. \ref{fig:fig_pdf_dcgen_Avg} for the DCGANs-based approach. Similarly, the market data for 2020 and 2021 are employed for training and testing, respectively. This approach does not perform well in generating negative prices. Statistics of generated ID average prices and the actual data are compared in Table \ref{tab:dcganstat}. Generated prices in the DCGANs-based approach have similar minimum, maximum, mean, and median (50\% percentile) compared to the actual data. Therefore, this approach is doing well in terms of statistic measures.

\begin{table}[h!]
	\caption{Statistic Summary of Average ID Prices in DCGANs-based Approach}\label{tab:dcganstat}
	\centering\!\!
	\begin{tabularx}{\columnwidth}{cYYYY}
		\toprule
		\rowcolor{color111}
		{\centering Statistic} &
		{\centering Actual} &
		{\centering Generated} \\ 
		\midrule
		\rowcolor{color110}
        mean 	&15.33 	&15.08 \\ \rowcolor{color112}
        standard deviation 	&17.37 	&24.26 \\ \rowcolor{color110}
        minimum 	&-29.21 &-29.21 \\ \rowcolor{color112}
        25\% 	&1.83 	&-3.30 \\ \rowcolor{color110}
        50\% 	&10.31 	&11.34 \\ \rowcolor{color112}
        75\% 	&22.50 	&30.25 \\ \rowcolor{color110}
        maximum 	&100.00 &100.00 \\ 
		\bottomrule
	\end{tabularx}\centering\\
\end{table}

\vspace{-5mm}
\subsection{No-U-Turn Sampler (NUTS)}\label{sec:results:nuts}
\begin{figure}[h!]
	\centering
 	\input{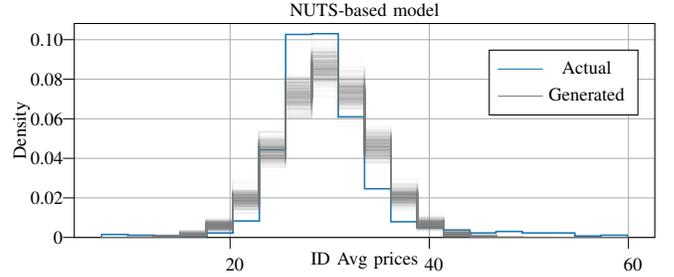} 
\vspace{-3mm}
	\caption{PDF of ID average price (EUR/MWh) for observed data (actual) and posterior predictive distributions (generated) in NUTS approach}
	\label{fig:fig_pdf_nuts_Avg}
\end{figure}

PDF of the ID average prices are shown in Fig. \ref{fig:fig_pdf_nuts_Avg} for the NUTS-based approach. This figures shows very good performance this approach in generating samples with the similar PDF as actual prices. Which is significantly better than the DCGANs approach.

All of the proposed approaches in this paper are practical depending on the type of study and it's context. For instance, in operational and short-term studies of power system with right amount of data, the LSTM-based approach is the best choice, as shown in Section \ref{sec:lstm} and Section \ref{sec:results:lstm}. On other hand, for more long-term studies with right amount of data, the DCGANs-based approach is a good choice as shown in Section \ref{sec:dcgans} and Section \ref{sec:results:dcgans}. Finally, for studies with limited access to data, the NUTS-based approach is the best choice as shown in Section \ref{sec:nuts} and Section \ref{sec:results:nuts}. Therefore, we suggest to the market participants to implement different ID price models. Then, depending on changes in the ID market regulations, choose the most precise model. For instance, after each changes in the market regulations, DCGANs-based and NUTS-based models are more accurate. Gradually more historical data is available with the updated regulations and the market participants may use the LSTM-based model. In addition, the market participants should always be updated with the upcoming changes in the market regulations.
\vspace{-5mm}
\subsection{Implementation}\label{sec:results:implementation}
LSTM and DCGANs algorithms are implemented in Keras
\cite{chollet2015keras} with backend of Tensorflow
\cite{abadi2015tensorflow}. NUTS algorithm is implemented in PyMC3
\cite{salvatier2016pymc3}. Source ID market data source is Nord Pool \cite{nordpool2021data}.
\vspace{-5mm}
\section{Conclusion}\label{sec:conclusion}
Uncertainties in electricity markets, e.g. Day-ahead (DA) and intraday (ID) markets, are increasing and they are continuously changing by introducing new regulations. In this paper, we have explored DA and ID markets by looking at the historical data and studied effect of time, area, bidding zones, and volume. It showed excessive and expanding importance of ID market price modeling. With the constant ID market changes, less historical data are available with the current ID market regulations. Therefore, a good price model for ID markets is crucial for the market participants. In the literature, there are not many practical models for electricity market price modeling, e.g. DA market price with sufficient data. We have proposed three approaches for price modeling based on long short term memory, deep convolutional generative adversarial networks, and No-U-Turn sampler which are all for ID market price modeling. To the best of our knowledge, this paper is the first in proposing several practical models for ID price modeling which is specially essential for ID markets. Finding the best approach depends on the participants theoretical knowledge about the distribution function of ID prices, their access to accurate and up-to-date data, and how valuable precise price is for the consumers. Our reflection is that the market participants need to employ a combination of the proposed price models and make operational decisions based on their context and the changes in the electricity markets.
\section{Nomenclature}\label{sec:appendix}
\printnomenclature

\ifCLASSOPTIONcaptionsoff
  \newpage
\fi
\bibliographystyle{IEEEtran}
\bibliography{IEEEabrv,bib}
\vfill

\end{document}